\documentclass[aps,pre,reprint,superscriptaddress,amsmath,amssymb,showpacs,floatfix]{revtex4-1}
\usepackage{graphicx}
\usepackage{bm}
\usepackage{color}
\usepackage{soul}

\begin{document}

\title{Non-equilibrium driven by an external torque in the presence of a magnetic field}

\author{Sangyun Lee}
\affiliation{Department of Physics, Korea Advanced Institute of Science and Technology, Daejeon 34051, Korea }
\author{Chulan Kwon}
\email{ckwon@mju.ac.kr}
\affiliation{Department of Physics, Myongji University, Yongin, Gyeonggi-Do, 17058, Korea }

\date{\today}

\begin{abstract}
We investigate a motion of a colloid in a harmonic trap driven out of equilibrium by an external non-conservative force producing a torque in the presence of a uniform magnetic field. We find that steady state exists only for a proper range of parameters such as mass, viscosity coefficient, and stiffness of the harmonic potential, and the magnetic field, which is not observed in the overdamped limit. We derive the existence condition for the steady state. We examine the combined influence of the non-conservative force and the magnetic field on non-equilibrium characteristics such as non-Boltzmann steady-state probability distribution function, probability currents, entropy production, position-velocity correlation, and violation of fluctuation-dissipation relation. 
\end{abstract}

\maketitle
 
\section{Introduction} 

Stochastic thermodynamics for the non-equilibrium motion of small systems has been an interesting issue since the discovery of the fluctuation theorem (FT). There have been many studies on non-equilibrium  fluctuation driven by external non-equilibrium sources such as non-conservative forces and time-dependent protocols which produce work and heat persistently~\cite{Evans:1993Prob,Evans:1994Equili,Gallavotti:1995Dynamical,Gallavotti:1995dyna,Jarzynski:1997Equili,Jarzynski:1997Nonequilibrium,Kurchan:1998Fluctuation,Crooks:1999Entropy,Lebowitz:1999Gallavotti,Hatano:2001Steady,Speck:2005Integral,Speck:2006restoring,Esposito:2010Three}. There have been extensive experimental studies, measuring work and confirming the FT~\cite{Wang:2002Experi,Hummer:2001Free,Liphardt:2002Equili,Collin:2005Verifi,Trepagnier:2004Experi,Garnier:2005Nonequili,Douarche:2006Work,Joubaud:2008Fluc,Hayashi:2010Fluc,Lee:2015Nonequili}. Under a particular circumstance, there exists a non-equilibrium steady state (NESS) characterized by non-Boltzmann distribution, non-zero current, non-zero rate of perpetual heat or work production, etc. 

The influence of magnetic field on non-equilibrium systems has been an interesting issue. Diffusion under no confining potential is an intrinsic non-equilibrium process and becomes more complicated under a magnetic field, observed in many plasmas. The diffusion under a magnetic field has been studied extensively~\cite{Taylor:1961diffusion,Kursunoglu:1963brownian,
Xiang:1993stochastic,Balescu:1997matter,Czopnik:2001brownian,
Jimenez:2007fokker}. Non-equilibrium system in a time-varying potential has also been studied in the presence of a constant magnetic field~\cite{Jayannavar:2007charged,Jimenez:2010work,Jimenez:2013brownian}.

Non-equilibrium driven by a non-conservative force in the presence of a magnetic field has not been considered in many places. The magnetic field does not produce any work so that the system does not undergo any energetic change solely due to the magnetic field. Contrary to deterministic dynamics, a usual circular motion cannot be observed due to thermal fluctuation in stochastic dynamics. The system under a conservative force in the presence of the magnetic field can reach a steady state with the Boltzmann distribution in the absence of any non-equilibrium source. Though the role of the magnetic field is not clear in this seemingly equilibrium situation, the dependence of the time-correlation functions on the magnetic field was found\cite{Balescu:1997matter} and will be examined more thoroughly in our study. Recently, it was reported that an unconventional entropy is produced by the magnetic field as generally done by a velocity-dependent force~\cite{Kwon:2016unconventional}.  It was also found that the proper overdamped limit cannot be found by neglecting an inertia term, but by investigating a colored noise induced by a magnetic field~\cite{Chun:2018emergence}. 

In our study, we investigate the motion of a charged colloid in a harmonic trap potential in the presence of a magnetic field, which is driven out of equilibrium by a torque-generating non-conservative force. We have investigated the overdamped limit in the absence of the magnetic field~\cite{Kwon:2011nonequilibrium,Noh:2013Multiple}.
In order to study the effect of the magnetic field rigorously, we investigate the motion in the phase space (position-velocity space).
In Sec.~\ref{model}, we present a mathematical setup for our model. In Sec.~\ref{sec:stability}, we derive the existence condition for a steady state. In Sec.~\ref{steady_state}, we find the probability distribution function (PDF) in NESS. As the characteristics of NESS, we find non-equilibrium probability currents in Sec.~\ref{NESS_current}, entropy production in Sec.~\ref{entropy_production}. In Sec.~\ref{correlation}, we derive two-time correlation functions among the pairs of positions and momenta. In Sec.~\ref{FDR}, we examine the violation of fluctuation-dissipation relation (FDR) caused by the non-conservative force and the magnetic field. We summarize our results in Sec.~\ref{summary}
\section{Model}
\label{model}
We consider a colloid of mass $m$ and charge $q_{\rm col}$ which is immersed in a two-dimensional liquid between parallel plates, as in an experimental setup.
We consider a Brownian motion under a harmonic potential mimicking an optical trap, which is driven out of equilibrium by a torque-generating non-conservative force. 
Let $\vec{r}=(x,y)$ and $\vec{v}=(v_x,v_y)$ be the position and velocity vectors of the colloid and $V(\vec{r})= (k_1x^2+k_2 y^2)/2$ be the trap potential with $k_1,~k_2>0$. We suppose that a uniform magnetic field $\vec{B}=B\hat{z}$ is applied perpendicular to the plane of the plates.
We consider an external force $\vec{f}_{\rm ex}=-\left(\begin{smallmatrix} 0& a_1\\a_2 &0\end{smallmatrix}\right)\cdot{\vec{r}}$ for $a_1\neq a_2$. It is a non-conservative force ($\vec{\nabla}\times\vec{f}_{\rm ex}\neq \vec{0}$) yielding a torque in $z$-direction and driving the colloidal motion out of equilibrium. 
Let $\gamma$ be a viscosity coefficient and $\beta$ be a fixed inverse temperature of the liquid. Under this condition, the motion of the colloid can be described by the Langevin equation written as $m \dot{\vec{v}}=-\vec{\nabla}V+
\vec{f}_{\rm ex} + q_{\rm col}B \vec{v}\times \hat{z}- \gamma \vec{v} + \vec{\eta}(t)$ where $\vec{\eta}(t)=(\eta_x(t),\eta_y(t))$ is a Gaussian noise vector with zero mean and variance given by $\langle\eta_i(t)\eta_i(t')\rangle=2\gamma  \beta^{-1}\delta_{ij}\delta(t-t')$ for $i, j=1,2$ denoting $x$, $y$. It can be rewritten as
\begin{equation}
    m \dot{\vec{v}} =-\mathsf{F}\cdot{\vec{r}} 
   - \mathsf{\Gamma} \cdot\vec{v}+ \vec{\eta}(t) 
\label{langevin}
\end{equation}
where $\mathsf{F}=\left(\!\begin{array}{cc} k_1& a_1\\a_2 &k_2\end{array}\right)$ and $\mathsf{\Gamma}=\left(\!\begin{array}{cc} \gamma & b\\-b&\gamma\end{array}\right)$ for $b=-q_{\rm col}B$. 

Let $\vec{q}=(x,y,v_x,v_y)$ be a state vector in the position-velocity space. Then, combining Eq.~(\ref{langevin}) and $\dot{\vec{r}}=\vec{v}$, we have the Langevin equation in extended dimensions as
\begin{eqnarray}
{\dot{ \vec{q} } }(t) = - \mathsf{M}\cdot {\vec{q}}(t) + {\vec{\xi}}(t)
\label{eq:underlangevin}\end{eqnarray}
where
\begin{equation}
\mathsf{M}= 
\left(
    \begin{array}{cc}
        \mathsf{0} &-\mathsf{I}   \\     
       \mathsf{F}/m & \mathsf{\Gamma}/m 
    \end{array}\right),
\end{equation}
where ${\vec{\xi}}(t)=( 0 , 0 , {\eta_x}(t)/m , {\eta_y}(t)/m)$. $\mathsf{0}$ and $\mathsf{I}$ are $2\times 2$ null and identity matrix, respectively. It belongs to the Ornstein-Ulenbeck process in four dimensions, which can be exactly solvable~\cite{Kwon:2005structure,Kwon:2011nonequilibrium}. 
The Fokker-Planck equation for the PDF $\rho(\vec{q},t)$ in the position-velocity space, called the Kramers equation, is written as
\begin{equation}
\partial_t\rho(\vec{q},t)=-\partial_{\vec{q}}\cdot(-\mathsf{M}\cdot\vec{q}
-\mathsf{D}\cdot\partial_{\vec{q}})\rho(\vec{q},t),
\label{FP_equation}
\end{equation}
where $\partial_t$ ($\partial_{\vec{q}}$) denotes partial differentiation with respect to $t$ ($\vec{q}$). 
$\mathsf{D}$ is a $4\times4$ diffusion matrix defined as $\gamma\beta^{-1}/m^2 
\left(
    \begin{array}{ll}
        \mathsf{0} &\mathsf{0}   \\     
       \mathsf{0} & \mathsf{I} 
    \end{array}\right)$.

When an initial PDF at $t=0$ is Gaussian, given as $\rho(\vec{q},0)\propto e^{- \vec{q} \cdot {\mathsf U}( 0 ) \cdot \vec{q}/2}$, the PDF at time $t$ can be written as
\begin{equation}
    \rho( \vec{q}(t) , t ) 
     = \left[\frac{\det\mathsf{U}(t)}{(2\pi)^4}\right]^{1/2} \exp\left[- \frac{1}{2} \vec{q}\cdot \mathsf{U}( t ) \cdot \vec{q}~\right] 
     \label{eq:PDFunction}
\end{equation}
where
\begin{equation}
    \mathsf{U}(t)^{-1}
    = \mathsf{U}^{-1}_{\rm ss} + e^{-\mathsf{M} t}\left[\mathsf{U}(0)^{-1}- \mathsf{U}_{\rm ss}^{-1}\right] e^{-\mathsf{M}^{\rm T} t }~.
\label{eq:PDF}
\end{equation}
Here, the superscript T denotes the transpose of a matrix.
$\mathsf{U}_{\rm ss}$ is the kernel of the steady state reached for $t\to\infty$. The formal expression for the steady state kernel is given by 
\begin{eqnarray} 
\mathsf{U}_{\rm ss}=\mathsf{(D+Q)}^{-1}\mathsf{M}~.
\label{steady_state}
\end{eqnarray} 
$\mathsf{Q}$ is an anti-symmetric matrix satisfying 
\begin{eqnarray}
\mathsf{QM+M^{\rm T}Q=DM-M^{\rm T}D}~.
\label{anti_symm_matrix}
\end{eqnarray}
Solving this equation for $\mathsf{Q}$, one can find the PDF~\eqref{eq:PDF} at time 
$t$~\cite{Kwon:2005structure,Kwon:2011nonequilibrium}.

\section{Existence of steady state }
\label{sec:stability}

The formula for the PDF in Eqs.~\eqref{eq:PDFunction}  and \eqref{eq:PDF} is meaningful only if $\mathsf{M}$ is positive-definite; otherwise, the steady state PDF does not exist.
The characteristic equation for the eigenvalue $\lambda$ of $\mathsf{M}$ is given as 
\begin{eqnarray}
0&=&\lambda^4 
- \frac{2 \gamma}{m} \lambda^3
+\frac{ b^2 + \gamma^2 + m ( k_1 + k_2 )}{m^2} \lambda^2\nonumber\\
&&- \frac{  b (a_1 - a_2 ) + \gamma (k_1 + k_2) }{m^2}\lambda
+\frac{ k_1 k_2 - a_1 a_2 }{ m^2 }
\label{eigenvalue_eq}
\end{eqnarray}
Then, existence condition for the steady state is given by the positivity of $\rm{Re}(\lambda)$, which guarantees the convergence of $\mathsf{U}(t)$ to $\mathsf{U}_{\rm ss}$ as $t$ increases, as seen in Eq.~(\ref{eq:PDF}).

\subsection{General criterion}
\subsubsection{In the absence of a magnetic field}

We first consider for zero magnetic field ($ b = 0 $). In this simple case, Eq.~(\ref{eigenvalue_eq}) can be solved as
\begin{equation}
\lambda = \frac{\gamma}{2m}
\left[1\pm\sqrt{1-\frac{2 m}{\gamma^2} \left( k_{1} + k_{2}\pm\sqrt{4 a_{1} a_{2}+(k_{1}-k_{2})^2} \right)}\right]~,
\label{eigenvalue_zeroB}
\end{equation}
and the other two eigenvalues are complex conjugates of theses.  
For brevity we write the eigenvalue in \eqref{eigenvalue_zeroB} as $\lambda=\gamma(1\pm\sqrt{\psi})/(2m)$. 
We find that the existence condition depends on the sign of $  4 a_1 a_2 + (k_1 - k_2 )^2 $. 

If $ 4 a_{1} a_{2}+(k_{1}-k_{2})^2 > 0$, $\psi$ is real. Then, the condition for ${\rm Re}(\lambda)>0$ is $\psi<1$, which leads to the existence condition 
\begin{equation}
    k_1 k _2 - a_1 a_2=\det \mathsf{F}>0~,
\label{1st_kind}
\end{equation}
where $\mathsf{F}$ is defined in Eq.~(\ref{langevin}).
The existence condition does not depend on mass $m$. Therefore, it can be applied to the overdamped limit for large $\gamma$ or small $m$, where Eq.~(\ref{langevin}) reduces to $\gamma\dot{\vec{r}}=-\mathsf{F}\cdot{r}+\vec{\eta}(t)$ in the position space. In this limit, $\det \mathsf{F}>0$ is nothing but the condition that $\vec{r}=\vec{0}$ be a stable fixed point. 

In the other case for $ 4 a_{1} a_{2}+(k_{1}-k_{2})^2 < 0 $, $\psi$ is imaginary.
We can write $\psi=L_1\pm iL_2$ where
$L_1=1-(2m/\gamma^2)(k_1+k_2)$, $L_2=(2m/\gamma^2)\sqrt{-4a_1a_2-(k_1-k_2)^2}$. Then, the condition that the smallest value of ${\rm Re}(\lambda)$ be positive can be found as $1-\sqrt{(\sqrt{L_1^2+L_2^2}+L_1)/2}>0$.
Then, we get the existence condition 
\begin{equation}
    k_1 + k_2+\frac{m}{ 2 \gamma^2 } [ ( k_1 - k_2 )^2 + 4 a_1 a_2 ] >0~.
    \label{2nd_kind}
\end{equation}
For a sufficiently small $m/\gamma^2$, the existence of the steady state is always guaranteed, hence this condition is beyond the overdamped limit. 

The 2-dimensional motion for $4a_{1}a_{2}+(k_1-k_2)^2>0$ can be shown to map to the previously studied cases such as a 2-dimensional motion subject to different noise sources (heat reservoirs) acting in the two perpendicular directions~\cite{Filliger:2007Brown}
and a one-dimensional model for two particles interacting via a harmonic force each of which is thermostatted to a different heat reservoir~\cite{Park:2016efficiency} . The latter is also equivalent to an electric circuit with two sub-circuits coupled via a capacitor~\cite{Ciliberto:2013Heat}. The stability criterion in Eq.~(\ref{1st_kind}) was examined for the heat engine designed from the former model~\cite{Park:2016efficiency}. Throughout the paper in the following, we consider the other case for $a_1a_2+(k_1-k_2)^2<0$, which is not derivable from the previous studies.
 
\subsubsection{In the presence of magnetic field}

The solution of the characteristic equation in Eq.~(\ref{eigenvalue_eq}) for nonzero $b$ can also be solved exactly with the help from Mathematica, but cannot be expressed in a simple form as Eq.~(\ref{eigenvalue_eq}). However, for small $b$, we can find the expression for the eigenvalues by using the perturbation expansion. Up to the first order in $b$, the correction to the zeroth order value $\lambda^{(0)}$ is found as 
\begin{equation}
\lambda^{(1)} = \frac{ ( a_{1} - a_{2} ) b }{ (  2 m \lambda^{(0)} - \gamma ) (2 m{\lambda^{(0)}}^2 - 2 \gamma \lambda^{(0)} + k_{1} + k_{2} )} \lambda^{(0)}~.
\label{eq:firstorder}
\end{equation} 
After some algebra, we find the positivity condition for the smallest value of ${\rm Re}(\lambda^{(0)}+\lambda^{(1)})$ as
\begin{eqnarray}
\lefteqn{1-\sqrt{\frac{\sqrt{L_1^2+L_2^2}+L_1}{2}}}\nonumber\\
&&+\frac{2m(a_1-a_2)b}{L_2\sqrt{L_1^2+L_2^2}}\sqrt{\frac{\sqrt{L_1^2+L_2^2}-L_1}{2}}>0~,
\end{eqnarray}
where $L_{1,2}$ are given in the last subsection. As a result, we have the existence condition for the steady state for a small $b$ as
\begin{equation}
k_1 +k_2 + \frac{m}{2\gamma^2} [ 4 a_{1}a_{2}+(k_1 - k_2)^2] +\frac{b (a_1 - a_2 )}{\gamma} > 0~,
\label{stability_perturbation}
\end{equation}
where $b$ is kept up to the first order.

\subsection{Isotropic case in the presence of a magnetic field}

We consider an isotropic case for $a_1=-a_2=a$, $k_1=k_2=k$, for which we can find the exact existence condition for steady state non-perturbatively for arbitrary $b$, while the condition for non-isotropic case can be found numerically. 
For the isotropic case, the eigenvalue equation in Eq.~(\ref{eigenvalue_eq}) reduces to
\begin{equation}
0=\lambda^4 
- \frac{2 \gamma}{m} \lambda^3
+\frac{ b^2 + \gamma^2 +2 mk}{m^2} \lambda^2
- \frac{  2ab + 2\gamma k}{m^2}\lambda
+\frac{k^2+ a^2 }{ m^2 }~.
\end{equation}
It is convenient to define dimensionless coefficients as follows:
\begin{equation}
A=\frac{ma}{\gamma^2}~,~~B=\frac{b}{\gamma}~,~~K=\frac{mk}{\gamma^2}.
\label{dimensionless}
\end{equation}
Then, the two typical eigenvalues of $\mathsf{M}$ can be written as
\begin{equation}
\lambda_{1,2}=\frac{\gamma}{2m}\left[1-iB \pm \sqrt{R}e^{i\phi/2}\right]
\label{eigenvalue}
\end{equation}
where 
\begin{eqnarray}
R&=&\sqrt{(1-B^2-4K)^2+(2B-4A)^2} , \nonumber\\
\phi&=&\tan^{-1}\frac{2B-4A}{1-B^2-4K}.
\label{euler}
\end{eqnarray}
The other two eigenvalues are complex conjugates of $\lambda_1$ and $\lambda_2$. 
Then, the condition ${\rm Re}(\lambda_{1,2})>0$ leads to $| \sqrt{R}\cos(\phi/2)|<1$, leading to 
\begin{equation}
1> \frac{R(1+\cos\phi)}{2}=\frac{R+1-B^2-4K}{2}.
\end{equation}
Simplifying it more, we find the stability condition as $K-AB+A^2>0$ or
\begin{equation}
\Omega=k+ab/\gamma-ma^2/\gamma^2>0~,
\label{stability}
\end{equation}
where we define $\Omega$ which frequently appears for other quantities obtained later. Note that it is consistent with Eq.~\eqref{stability_perturbation} in the isotropic limit. It implies that all the higher-order corrections in $b$ to Eq.~\eqref{stability_perturbation} vanishes in the isotropic limit, which is non-trivial to show rigorously in the perturbation scheme. 

\begin{figure}
\centering
\includegraphics[width=0.3\textwidth]{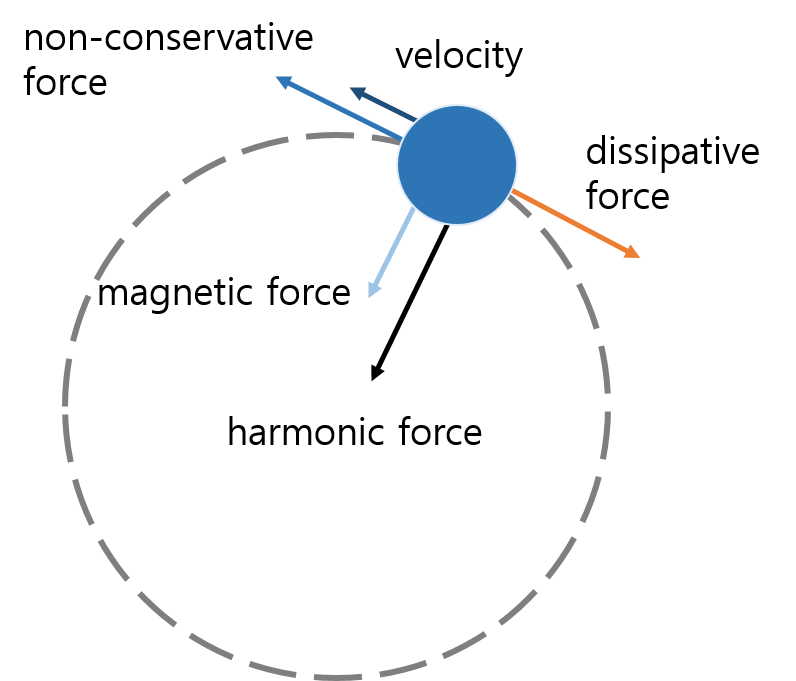}
\caption{\label{fig:critical} A circular orbit and involved forces. The dissipative force is $-\gamma\vec{v}$. The figure is drawn for $a,~b>0$. The harmonic force and magnetic force for $ab>0$ ($ab<0$) are in the same (opposite) direction so that they strengthen (weaken) centripetal force. }
\end{figure}

We provide a more physical derivation based on the stability of a fixed point. A deterministic trajectory of the motion generated by Eq. (\ref{eq:underlangevin}) is given by $\dot{\vec{q}}_{\rm d}=-\mathsf{M}\cdot\vec{q}_{\rm d}$. In polar coordinates $(r,\theta)$, there is a fixed point at $r=0$, which is either stable or unstable in the parameter space $(m, \gamma, a, b, k)$. At the critical boundary in the parameter space, there exists a fixed circular orbit the radius of which depending depends on an initial condition and hence infinitely many circular orbits including $r=0$. A circular orbit satisfies the two force-balance equations in radial and angular directions, given as  $mr{\dot\theta}^2=kr+br\dot\theta$ and $mr\ddot{\theta}=-\gamma r\dot\theta+ar=0$. Eliminating $\dot\theta$, we find $mr(a/\gamma)^2=(k+ab/\gamma)r$ where the right-hand-side is the centripetal force for the circular orbit, hence $\Omega=0$ from Eq.~(\ref{stability}).  For $\Omega>0$, a deterministic trajectory converges to $r=0$ as time evolves, which comes up with a stable PDF through fluctuation by noise. For $\Omega<0$, however, any trajectory diverges to $r=\infty$ so that noise cannot produce any stable PDF. Figure~\ref{fig:critical} shows a circular orbit where harmonic and magnetic forces in radial direction. For $ab>0$, the two forces are in the same radial direction so as to strengthen centripetal force, and vice versa for $ab<0$.

\begin{figure}
\centering
\includegraphics[width=0.4\textwidth]{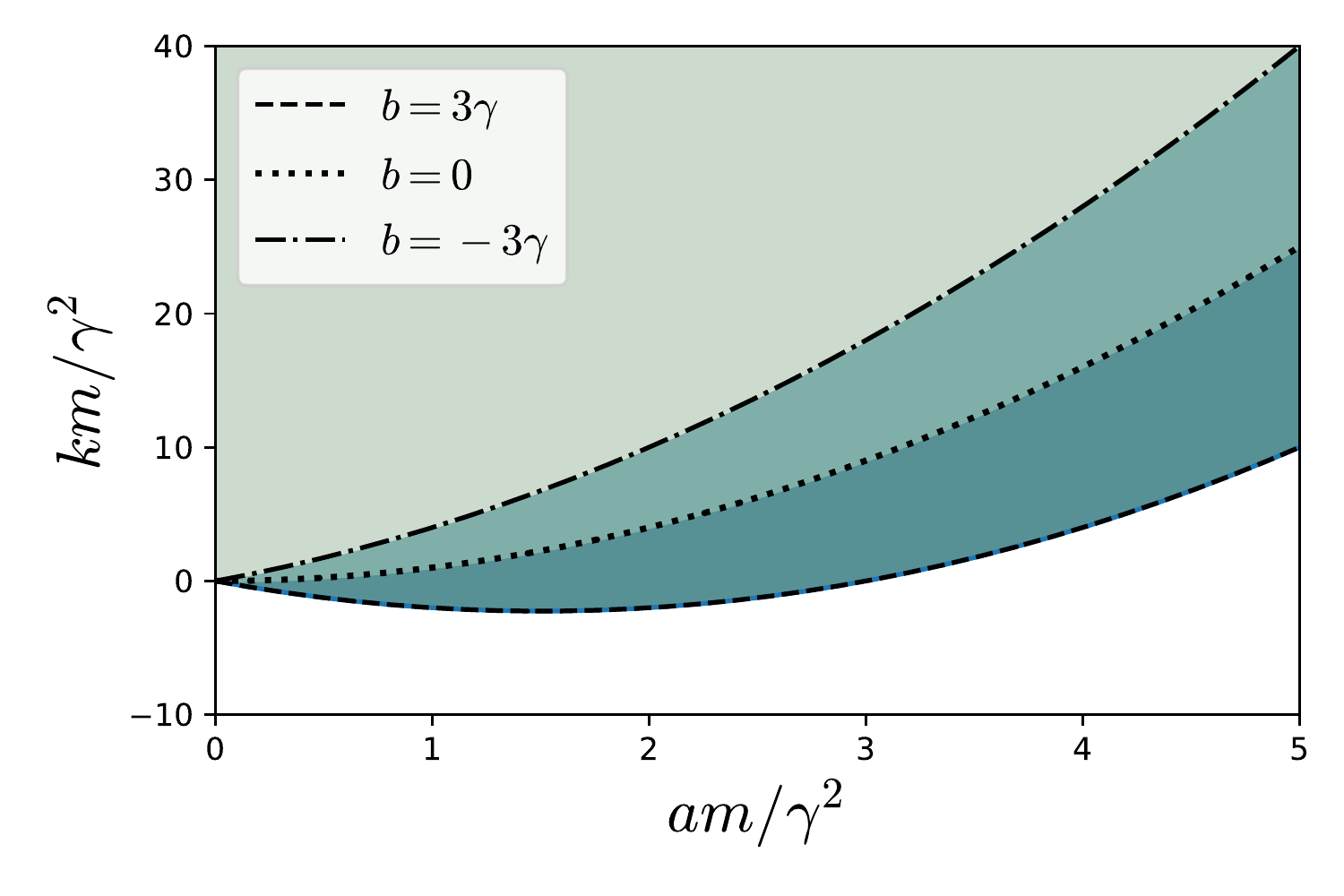}
\caption{\label{fig:stability3} The existence region for steady state in a parameter space for $k$ and $a>0$. The stable region is above the boundary line. Boundaries are drawn for $b = -3 \gamma ,\, 0 ,\, 3\gamma$.
The larger $b$ and the smaller $a$, the more widened the stable region. }
\end{figure}

The external torque gives an acceleration in angular direction to drive a spiral motion outward from the origin, so it tends to depress the stability, as seen from the last term, $-ma^2/\gamma^2$ in Eq.~\eqref{stability}. For $ab>0$, the magnetic field is in the same direction as the torque $\nabla\times \vec{f}_{\rm ex}$ so that it yields a magnetic force in the same centripetal direction as the harmonic force, and vice versa for $ab<0$. Therefore, the magnetic field tends to enhance (depress) the stability for $ab>0$ ($ab<0$), as seen in the second term, $ab/\gamma$, in Eq.~\eqref{stability}.  Figure~\ref{fig:stability3} shows the diagram for the existence of steady state in $k$-$a$ space for various values of $b$, where
the competing and supplementary tendencies in the influence of $a$ and $b$ on the stability condition is well observed.
 
\section{Non-equilibrium Steady state}
\label{steady state}

In the existence region satisfying the condition $\Omega >0$ in \eqref{stability}, we can find the steady state PDF  and show it explicitly for the isotropic case. First, we solve  Eq.~(\ref{anti_symm_matrix}) for the anti-symmetric matrix $\mathsf{Q}$, which can be converted into a set of linear equations for six unknown elements of the matrix. We find 
\begin{equation}
\mathsf{Q}=\frac{1}{\Omega}\left(
\begin{array}{cccc}
0&\frac{a}{\gamma}& -\frac{ab+\gamma k}{m\gamma}&0\\
-\frac{a}{\gamma}&0&0&-\frac{ab+\gamma k}{m\gamma}\\
\frac{ab+\gamma k}{m\gamma}&0&0&\frac{ab^2+\gamma b k+akm}{m^2\gamma}\\
0& \frac{ab+\gamma k}{m\gamma}&-\frac{ab^2+\gamma b k+akm}{m^2\gamma}&0
\end{array}
\right)~.
\end{equation}
From Eq.~(\ref{steady_state}), we have
\begin{eqnarray}
\mathsf{U}_{\rm ss}=\beta
\left(
\begin{array}{cccc}
 \frac{a b + \gamma k}{\gamma } & 0 & 0 & -\frac{ a m }{ \gamma } \\
 0 & \frac{a b + \gamma k}{\gamma } & \frac{ a m }{ \gamma } & 0 \\
 0 & \frac{a m}{ \gamma } & m & 0 \\
 -\frac{a m}{\gamma } & 0 & 0 & m \\
\end{array}
\right)
\label{steady_state_sol}
\end{eqnarray}

For $a=0$, $\mathsf{U}_{\rm ss}$ is equal to that for the equilibrium Boltzmann PDF, independent of a magnetic field. It is well explained from the fact that the magnetic field does not work. However, for the transient period for $t<\infty$, a relaxation behavior of the PDF in time towards the Boltzmann PDF is determined by $e^{-\mathsf{M}t}$ and $e^{-\mathsf{M}^{\rm T}t}$, as seen in~\eqref{eq:PDF}, and hence various forms of  exponential decaying with sinusoidal oscillation as $e^{-(\lambda_i+\lambda_j) t}$ for all possible $i,~j$. As seen in Eq.~(\ref{eigenvalue}), even for $a=0$, eigenvalues $\lambda_i$'s depend on $b$, so the transient PDF depends on $b$.  

For $a\neq 0$ in non-equilibrium, the steady state PDF ($\rho_{\rm ss}$) depends on $b$ as well as the transient one. We can observe that positions and velocities are coupled in the PDF, as seen from the off-diagonal elements of $\mathsf{U}_{\rm ss}$, which gives rise to a non-Maxwellian distribution as an important characteristics of non-equilibrium steady state (NESS). One can observe $\beta^{-1}\ln\rho_{\rm ss}=-(m/2) \left[(v_x+ay/\gamma)^2+(v_y-ax/\gamma)^2\right]+\cdots$. Then, we have a nonzero average velocity at a fixed position, given as
\begin{equation}
\langle\vec{v}\rangle_{\vec{v}}=-\frac{a}{\gamma}\mathsf{A}\cdot\vec{r}~,
\label{ave_velocity}
\end{equation}
where $\langle\, \cdots\, \rangle_{\vec{v}}=\int d\vec{v}\rho_{\rm ss}(\vec{r},\vec{v})(\cdots)$ denotes the average of the given quantity over $\vec{v}$ for a fixed position. $\mathsf{A}=\left(\begin{array}{ll} 0 & 1\\-1 & 0\end{array}\right)$ is an anti-symmetric matrix. It manifests a nonzero  probability current, also known as an important property of NESS. This property is more examined in the next section. 

We can find second moments in the steady state as
\begin{equation}
\langle \vec{q}\vec{q}\rangle=\mathsf{U}_{\rm ss}^{-1}=\frac{1}{\beta \Omega}\left(
\begin{array}{cccc} 1&0&0&\frac{a}{\gamma}\\ 0&1&-\frac{a}{\gamma} & 0\\0&-\frac{a}{\gamma}&\frac{ab+k\gamma}{m\gamma}& 0\\ \frac{a}{\gamma} & 0&0&\frac{ab+k\gamma}{m\gamma}
\end{array}\right)~,
\label{equal_time_corr}
\end{equation}
where $(\vec{q}\vec{q})_{ij}=q_iq_j$ is a $4\times 4$ dyad (outer product) of a  state vector in the position-velocity space.

\section{Non-equilibrium probability currents}
\label{NESS_current}

NESS is characterized by a nonzero irreversible current in the variable space. We follow a well-established formalism in the textbook by H. Risken~\cite{Risken:1996fokker}. 
The Fokker-Planck equation \eqref{FP_equation} can be rewritten as
\begin{equation}
\partial_t\rho(\vec{q},t)=-\partial_{q_i}(f_i(\vec{q})-D_{ij}\partial_{q_j})\rho(\vec{q},t).
\end{equation}
Parity $\epsilon_i$ in time reversal: $t\to \tau-t$ for a final time $\tau$ is either $+1$ for position coordinates ($i=1,2$) or $-1$ for velocity coordinates ($i=3,4$). 
Then, the drift terms $f_i$'s are decomposed into reversible and irreversible parts
as
\begin{equation}
f^{\rm rev}_i=\frac{f_i(q_j)-\epsilon_i f_i(\epsilon_jq_j)}{2}~,~
f^{\rm irr}_i=\frac{f_i(q_j)+\epsilon_i f_i(\epsilon_jq_j)}{2}.
\end{equation}
The Fokker-Planck equation can be written as $\partial_t\rho=-\partial_{\vec{q}}\cdot\vec{j}_{\vec{q}}$ in terms of the probability current $\vec{j}_{\vec{q}}$, which can also be decomposed into
the reversible and irreversible parts as
\begin{equation}
 j_i^{\rm rev}=f_i^{\rm rev}\rho~,~j_i^{\rm irr}=(f_i^{\rm irr}-D_{ij}\partial_{q_j})\rho ~.
\end{equation}
Note that $j^{\rm irr}_i$ exists only in the velocity space, i.e., $j=3,4$. We use $\vec{j}_{\vec{r}}=(j_1,j_2)$ and $\vec{j}_{\vec{v}}=(j_3,j_4)$. In a usual convention, the magnetic field is to flip ($\vec{B}\to -\vec{B}$) in time reversal. In this study, however, we use a different rule without flipping $\vec{B}$ in time reversal in order to investigate irreversibility in dynamics under a given magnetic field. Then, we have
\begin{eqnarray}
\vec{j}^{\rm rev}_{\vec{r}}&=&\vec{v}\rho~,~~\vec{j}^{\rm rev}_{\vec{v}}=-\frac{1}{m}\mathsf{F}\cdot\vec{r}\rho\nonumber\\
\vec{j}^{\rm irr}_{\vec{v}}&=&\left[-\frac{1}{m}\Gamma\cdot\vec{v}-\mathsf{D}_{\rm red}\cdot\partial_{\vec{v}} \right] \rho~,
\label{eq:currents}
\end{eqnarray}
where $\mathsf{D}_{\rm red}=(\gamma\beta^{-1}/m^2)\mathsf{I}$ for $2\times 2$ identity matrix $\mathsf{I}$. 

The current in the velocity space, $\vec{j}_{\vec{v}}=\vec{j}_{\vec{v}}^{\rm rev}+\vec{j}_{\vec{v}}^{\rm irr}$, is the sum of forces per mass times PDF. We call $-m\mathsf{D}_{\rm red}\cdot\partial_{\vec{v}}\ln \rho$ stochastic force, which originates from noise in the Langevin dynamics. As seen in Eq.~\eqref{eq:currents}, any position-dependent force belongs to $\vec{j}^{\rm rev}_{\vec{v}}$. On the other hand, the dissipative force ($-\gamma\vec{v}$), the stochastic force, and the magnetic force  belong to $\vec{j}_{\vec{v}}^{\rm irr}$. The dissipative and stochastic forces in $\vec{j}^{\rm irr}_{\vec{v}}$ contribute to the production of heat, which is consistent with the definition of heat production rate in the system: $[-\gamma\vec{v}+\vec{\eta}(t)]\circ\vec{v}$ for $\circ$ denoting the Stratonovich convention. The role of the magnetic force in the irreversible current is intriguing because  it costs no energy, which will be discussed in this and the following section.

In the steady state, we find the irreversible current by using Eq.~(\ref{steady_state_sol}) as
\begin{equation}
\vec{j}_{\vec{v}}^{\rm irr}=\left[\frac{a}{m}\mathsf{A}\cdot\vec{r}-\frac{b}{m}\mathsf{A}\cdot\vec{v}~\right]\rho_{\rm ss}~,
\label{j_irr}
\end{equation}
The first term in this equation is exactly equal to minus the non-conservative force per mass in $\vec{j}^{\rm rev}_{\vec{v}}$. This means that the heat produced by this force exactly cancels the work produced by the non-conservative force, so the system can stay in the steady state. The total remaining force is given as 
\begin{equation}
m\rho_{\rm ss}^{-1}(\vec{j}_{\vec{v}}^{\rm rev}+\vec{j}_{\vec{v}}^{\rm irr})=-k\vec{r}-b\mathsf{A}\vec{v}~.
\label{remaining_current}
\end{equation} 
On the other hand, the reversible current $\vec{j}_{\vec{r}}^{\rm rev}=\vec{v}\rho_{\rm ss}$ in the position space is random in $\vec{v}$. We find the average current in position space as 
\begin{equation}
\langle\vec{j}^{\rm rev}_{\vec{r}}\rangle_{\vec{v}}=\int {\rm d}\vec{v} ~\vec{v}\rho_{\rm ss}(\vec{q})
=-\frac{a}{\gamma}\mathsf{A}\cdot\vec{r}~\widetilde{\rho}_{\rm ss}(\vec{r})
\label{circulation}
\end{equation}
where $\widetilde{\rho}_{\rm ss}(\vec{r})=[\beta\Omega/(2\pi)]e^{-\beta\Omega r^2/2}$ is the reduced steady-state PDF for $\vec{r}$. This average current circulates in the position space and the remaining current in the velocity space in Eq.~\eqref{remaining_current} provides a centripetal force necessary for such circulation. For a more rigorous proof, we write the PDF in polar coordinates as 
\begin{eqnarray}
\lefteqn{\rho_{\rm ss}(v_r,v_\theta,r,\theta)=\frac{\beta^2m\Omega}{(2\pi)^2}} \nonumber\\
&&\times\exp\!\!\left[\!-\frac{\beta m}{2}\!\!\left[v_r^2+\left(v_\theta-\frac{a}{\gamma}r\right)^2\right]\!\!-\frac{\beta\Omega}{2} r^2\right]~.
\label{PDF_polar}
\end{eqnarray}
The existence of the average circular current requires the condition: $\langle k\vec{r}+b\mathsf{A}\cdot\vec{v}\rangle=\langle mv_\theta^2 \hat{r}/r\rangle$.
The l.h.s and r.h.s of this condition are found as $\langle (k+ab/\gamma)r\rangle\hat{r}$ and 
$\langle  1/(\beta r) + (ma^2/\gamma^2)r\rangle\hat{r}$, respectively.
The two sides are found to be the same by using $\langle 1/r\rangle= \sqrt{\beta\Omega\pi/2}$ and $\langle r\rangle=\sqrt{\pi/(2\beta\Omega)}$ given from Eq.~(\ref{PDF_polar}). The magnetic field is shown to be a source for the circulating  current in the position space in addition to the torque-generating non-conservative force. It is interesting that the circular current could be possible even for $k=0$ if $ab>0$, rigorously for $ab>ma^2/\gamma$ ($\Omega>0$). 

The detailed balance (DB) characterizes dynamical reversibility, for which the condition is given as
\begin{equation}
\langle\vec{q'}|e^{-{\cal H}_{\rm FP}\Delta t}|\vec{q}\rangle\rho_{\rm ss}(\vec{q})=\langle\mathsf{\epsilon}\vec{q}|e^{-{\cal H}_{\rm FP}\Delta t}|\mathsf{\epsilon}\vec{q'}\rangle\rho_{\rm ss}(\mathsf{\epsilon}\vec{q'})~,
\end{equation}
where $(\mathsf{\epsilon}\vec{q})_{i}=\epsilon_i q_i$, ${\cal H}_{\rm FP}=\partial_{q_i}(f_i(\vec{q})-D_{ij}\partial_{q_j})$ is a non-Hermitiain Fokker-Planck operator, and $\Delta t$ is the time taken for the transition between the two states. It is shown~\cite{Risken:1996fokker, Lee:2013Fluc,Kwon:2016unconventional} that the DB holds only if 
\begin{equation}
\rho_{\rm ss}(\vec{q})=\rho_{\rm ss}(\mathsf{\epsilon}\vec{q})~,~ \vec{j}^{\rm irr}_{\vec{v}}=\vec{0}~
\label{DB}
\end{equation}
In our case, the DB is found to be broken. First, we clearly see $\rho_{\rm ss}(\vec{r},\vec{v})\neq \rho_{\rm ss}(\vec{r},-\vec{v})$ from position-velocity coupling in Eq.~(\ref{steady_state_sol}). Second, we find a nonzero irreversible current in Eq.~(\ref{j_irr}). The magnetic field as a part of the irreversible current is partly  responsible for the dynamical irreversibility manifested by the circulation in the position space besides its own contribution to the irreversibility in the velocity space.

\section{Entropy production}
\label{entropy_production} 
The total entropy $\Delta S$ produced for $0<t\le t_f$ in the system and bath can be regarded as a quantity to measure dynamic irreversibility. It is known to be found from the ratio of two path probabilities, given as
\begin{equation}
\Delta S=\ln\frac{\rho(\vec{q}_0,0)\Pi[\vec{q}(t)|\vec{q}_0;\lambda(t)]}{\rho(\vec{q}_{f},t_f)\Pi[\mathsf{\epsilon}\vec{q}(t_f-t)|\mathsf{\epsilon}\vec{q}_{f};\lambda(t_f-t)]}~,
\label{path_ratio}
\end{equation} 
where $\Pi[\vec{q}(t)|\vec{q}_0;\lambda(t)]$ ($\Pi[\epsilon\vec{q}(t_f -t)|\mathsf{\epsilon}\vec{q}_f;\lambda(t_f-t)]$) is the conditional probability of the system evolving along a path $\vec{q}(t)$ (time-reverse path $\epsilon\vec{q}(t_f-t)$) for given $\vec{q}_0$ ($\mathsf{\epsilon}\vec{q}_f$) at $t=0$ for $0<t\le t_f$. $\lambda(t)$ is a time-dependent protocol not considered in our study. $\Delta S$ satisfies the fluctuation theorem: $\langle e^{-\Delta S}\rangle=1$~\cite{Speck:2005Integral,Speck:2006restoring,Esposito:2010Three} and has a non-negative average as a corollary: $\langle\Delta S\rangle\ge0$. 
In the the absence of any velocity-dependent force, $\Delta S$ turns out to be equal to the sum of the Shannon entropy change, $-\Delta \ln \rho$, and the dissipated heat production $Q$ divided by temperature. Then, the dynamical irreversiblity accompanies energetic irreversibility in heat production. In particular, the two kinds of irreversibility are equivalent in the steady state with no Shannon entropy change.

In the presence of a velocity-dependent force, however, $\Delta S$ is found to have an unconventional contribution, $\Delta S_{\rm uc}$, resulting in a modified expression $\Delta S=-\Delta\ln \rho+Q/T+\Delta S_{\rm uc}$~\cite{Kwon:2016unconventional}. Various types of velocity-dependent forces have been considered in active matters~\cite{Marchetti:2013Hydro,Kim:2004Entropy,Kim:2007Fluc, Schweitzer:2007brownian,Romanczuk:2012Active,Ganguly:2013Stochastic,Chaudhuri:2014Active,Courty:2001Quantum,G:2007Tuning}
and a magnetic force is the only natural one.
The rate of the entropy production is given as
\begin{equation}
\dot{S}=-\frac{d}{dt}\ln\rho + \frac{\dot{Q}}{T}  + \dot{S}_{\rm uc}
\end{equation}
where 
\begin{eqnarray}
\dot{Q}&=&\dot{W}-\frac{d E}{dt}=-a\vec{v}\cdot\mathsf{A}\cdot\vec{r}-\frac{d}{dt}\!\!\left(\!\frac{m\vec{v}^2}{2}+\frac{k\vec{r}^2}{2}\!\right),\\
\dot{S}_{uc}&=&\frac{m}{\gamma}(\vec{f}^{\rm irr}+\gamma\vec{v})\cdot(\vec{f}^{\rm irr}-\gamma\vec{v})\nonumber\\
&&-\frac{1}{m}\partial_{\vec{v}}\cdot(\vec{f}^{\rm rev}-\vec{f}^{\rm irr}-\gamma\vec{v})~,
\end{eqnarray}
where $\dot{W}$ is the rate of work done by the non-conservative force. In obtaining $\dot{S}_{\rm uc}$ from Eq.~\eqref{path_ratio}, we change the sign of velocity in a time-reverse path, but not the protocol (coefficient) for the velocity-dependent force for the purpose to investigate the irreversibility under a fixed protocol. In fact, we do fix the direction of $\vec{B}$ in a time-reverse path. We are interested in a local irreversibility of the system under a fixed protocol provided from an external agent.
 
From the previous study~~\cite{Speck:2006restoring,Esposito:2010Three,Kwon:2016unconventional}, we have
\begin{equation}
\langle\dot{S}\rangle=\int d\vec{q}~\frac{\vec{j}^{\rm irr}_{\vec{v}}\cdot\mathsf{D}^{-1}_{\rm red}\cdot\vec{j}^{\rm irr}_{\vec{v}}}{\rho}\ge 0,
\label{EP}
\end{equation}
which is certainly non-negative. It explicitly shows the second law of thermodynamics in the presence of a velocity-dependent (magnetic) force. Interestingly, only the irreversible current contributes to the irreversibility appearing in a non-equilibrium process. 
 
For our case, $\vec{f}^{\rm irr}=-\Gamma\cdot\vec{v}/m$ and $\vec{f}^{\rm rev}=-\mathsf{F}\cdot\vec{r}/m$. We find $\dot{S}_{\rm uc}=(\beta b^2/\gamma){v}^2$, which is non-zero even when there is no non-conservative force. 
In the steady state, we find the average values of the components of $\dot{S}$ by using Eq.~\eqref{equal_time_corr} as
\begin{eqnarray}
\beta \langle\dot{Q}\rangle&=&-\beta a \langle v_xy-v_yx\rangle=\frac{2a^2}{\gamma\Omega}~,
\label{heat}\\
\langle\dot{S}_{\rm uc}\rangle&=&\frac{\beta b^2}{\gamma}\langle v_x^2+v_y^2\rangle=\frac{2b^2}{m\gamma\Omega}\left(k+\frac{ab}{\gamma}\right)~.
\label{EP_uc}
\end{eqnarray}
The total irreversibility quantified by $\langle\dot{S}\rangle$ has contributions from the two components, the non-conservative and the magnetic force, as seen in the irreversible current in Eq.~\eqref{j_irr}. The heat dissipation rate in Eq.~\eqref{heat} has the contribution from the first component and the unconventional entropy production rate in Eq.~\eqref{EP_uc} has the combined contribution from the both components, as seen from the dependence on $b^2$ and $ab$, respectively. Note that the magnetic force can have influence on the circulation current in the position space only by being accompanied by the non-conservative force. For $a=0$, there is no such circulation and heat production, but the irreversibility due to helicity, which is a tendency of circulation, is still present, which is measured by $\langle \dot{S}_{\rm uc}\rangle$.

\section{Two-time Correlation functions} 
\label{correlation}

Correlation functions between  position and velocity coordinates at different times are found by using the formula~\cite{Kwon:2011nonequilibrium}, given as
\begin{equation}
\mathsf{C}(t,t')\!=\!\langle\vec{q}(t) \vec{q}(t')\rangle =
\begin{cases}
e^{-(t-t')\mathsf{M}} \mathsf{U}^{-1}(t')~,& t > t' \\
 \mathsf{U}^{-1}(t') e^{-(t'-t)\mathsf{M}^{\rm T}} ~,& t < t'
\end{cases}
\label{correlation_matrix}
\end{equation}
where $\mathsf{U}(t)$ is the kernel for the PDF at time $t$, given in Eq.~(\ref{eq:PDF}).
We consider the correlation functions in the steady state, so $\mathsf{U}(t)=\mathsf{U}_{\rm ss}$. As a result, the two-time correlation functions only depend on the difference of two times. The equal-time correlation functions are found from $\mathsf{U}_{\rm ss}^{-1}$ in Eq.~(\ref{equal_time_corr}).

For the isotropic case, rotational symmetry yields
\begin{equation}
\langle x(t)x(t')\rangle =\langle y(t)y(t')\rangle,~
\langle x(t)y(t')\rangle = - \langle y(t)x(t')\rangle.
\label{eq:isorotsymm}
\end{equation}
Finding $\langle x(t)x(t')\rangle$\ and $\langle x(t)y(t')\rangle$, all other correlation functions can be generated by differentiating with respect to one of two times or by exchanging $x$ and $y$ components  with minus sign. For example, $ \partial_t \langle x(t) y(t')\rangle= \langle v_x(t) y(t')\rangle =-  \langle v_y(t) x(t')\rangle$. 

For $\tau=t-t'$, we write
\begin{equation}
e^{-\mathsf{M}\tau}=\sum_{i=1}^4 e^{-\lambda_i \tau}|i\rangle\langle i|
\end{equation}
where $|i\rangle$ ($\langle i|$) is an orthonormal right (left) eigenvector for an eigenvalue $\lambda_i$ for $\mathsf{M}$, i.e., $\langle i|j\rangle=\delta_{ij}$. Using the definition of $R$ and $\phi$ in Eq.~(\ref{euler}), the two kinds of time-dependent terms in $e^{-\mathsf{M}t}$ are found as
\begin{eqnarray}
c_i(\tau) &=& e^{-{\rm Re}(\lambda_i)\tau} \cos[{\rm Im}(\lambda_i)\tau+ \phi/2 ]~,\\
s_i(\tau)  &=& e^{-{\rm Re}(\lambda_i)\tau} \sin[{\rm Im}(\lambda_i)\tau+ \phi/2 ]~,
\label{oscillatory}
\end{eqnarray}
where $\lambda_i$'s for $i=1,2$ are the two typical eigenvalues in Eq.~(\ref{eigenvalue}).
Writing $\mathsf{C}(t,t')=\mathsf{C}(\tau)$, we have
\begin{eqnarray}
 C_{xx}(\tau)
     &=& 
  \frac{1}{\beta \Omega}  \sum_{ i = 1, 2}
   [\alpha_i c_i(\tau)+\beta_i s_i(\tau) ],~\nonumber\\
C_{xy}(\tau)
    &=& \frac{1}{\beta \Omega}
    \sum_{ i = 1, 2}
    [ \alpha_i s_i(\tau) - \beta_i c_i(\tau) ]~,
\label{eq:twocorr}
\end{eqnarray}
with
\begin{eqnarray}
    \alpha_{1,2} &=& \frac{1}{2}\left[\cos(\phi/2)\pm\frac{1}{\sqrt{R}}\right],\nonumber\\ 
    \beta_{1,2} &=& \frac{1}{2}\left[ \sin(\phi/2) \pm \frac{2 A - B}{\sqrt{R}} \right], 
\end{eqnarray}
where the upper (lower) sign is for the subscript $1$ ($2$). 
The parameters used are defined in Eqs.~(\ref{dimensionless}) and (\ref{euler}).
The other correlation functions derivable from Eq.~(\ref{eq:twocorr}) are given in Appendix~\ref{appendix:correl}.

\begin{figure}
\centering
\includegraphics[width=0.45\textwidth]{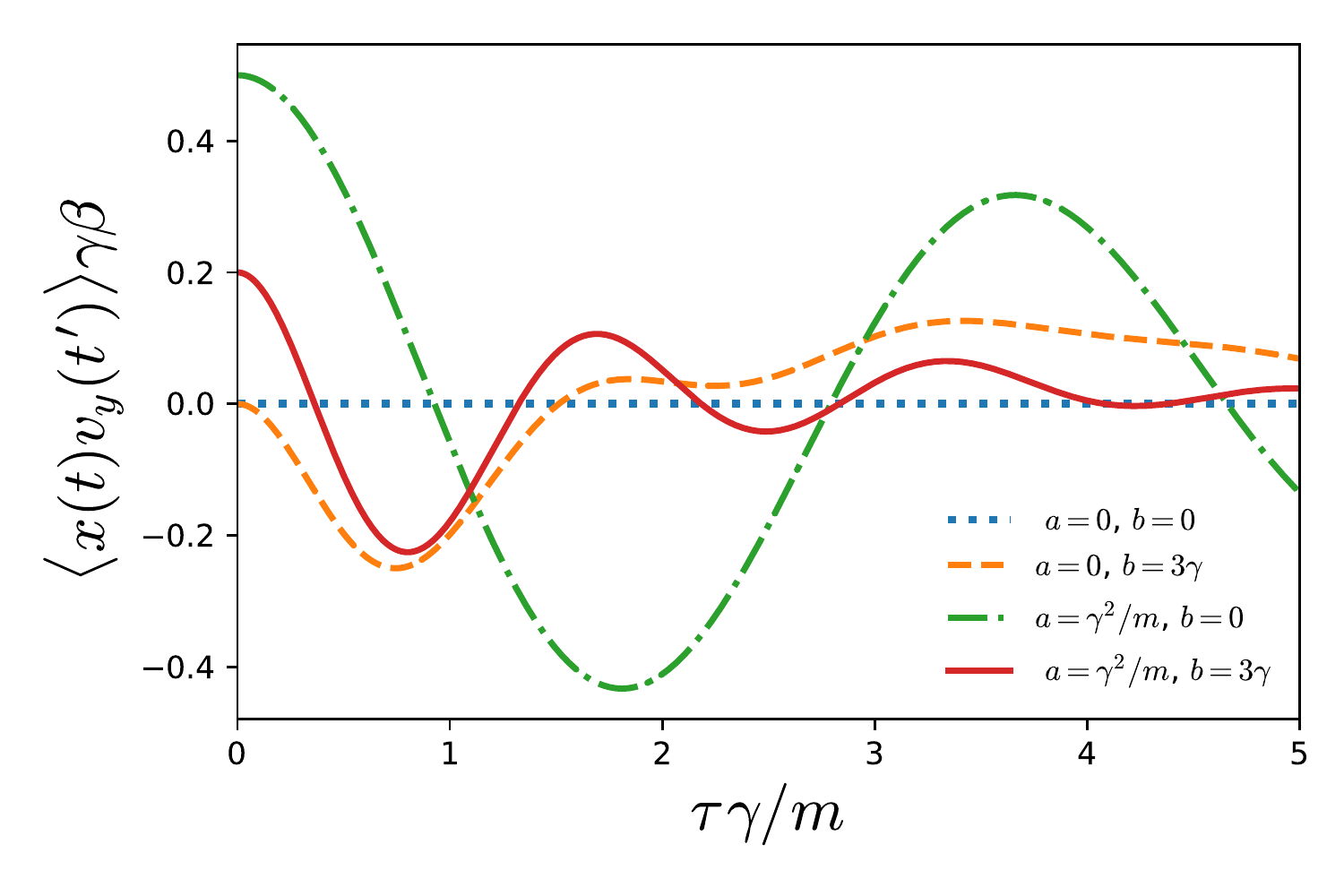}
\caption{\label{fig:twocorrelation} The plot of $ C_{xv_y}(\tau)$ for various $a,~ b>0$ with fixed $k = 3 \gamma^2 /m $. There is a non-vanishing correlation for $b\neq 0$ and $a=0$, which is distinguishable from normal equilibrium. The amplitude of the correlation decreases as $b$ increases. }
\end{figure}

There is circulating probability current in the position space. It is manifested in a strong correlation between position and velocity in perpendicular directions to each other. We plot a correlation function $ C_{xv_y}(\tau)=\langle x(\tau)v_y(0)\rangle$ in Figure~\ref{fig:twocorrelation}. It is interesting that it is nonzero even for $a=0$ and $b\neq 0$, which signals a tendency of helicity around the direction of the magnetic field. Interestingly, all the correlation functions have the same factor $\Omega$ in the denominator. Therefore, the nearer is the parameter set from the existence boundary (the larger $\Omega$), the smaller is the amplitude of the correlation function. In Figure~\ref{fig:twocorrelation} drawn for $a,~b>0$, the correlation function for $b=0$ has a larger amplitude than that for $b>0$ with a larger value of $\Omega$.

\section{Violation of fluctuation-dissipation relation} 
\label{FDR}

The fluctuation-dissipation relation (FDR) is known to hold for equilibrium. Recently, the violation of FDR was found to be related with the heat produced during a non-equilibrium process~\cite{Harada:2005equality,Harada:2006energy}. The FDR was found to hold in the presence of a magnetic field where there is no non-equilibrium source to produce heat~\cite{Lee:2017nonequilibrium}. We examine the FDR in our case where both a non-conservative and a magnetic force are present.

Under an arbitrarily small perturbative force $\vec{h}(t)$, the Lagenvin equation in Eq.~\eqref{langevin} is written as $m\dot{\vec{v}} =-\mathsf{F}\cdot{\vec{r}} - \mathsf{\Gamma} \cdot\vec{v} +\vec{h}(t)+ \vec{\eta}(t)$. The response function for $ \langle \vec{q}(t) \rangle $ with respect to variation $\delta\vec{h}(t')$ is defined as 
\begin{equation}
\mathsf{R}(t,t')=\left.\frac{\delta}{\delta \vec{h}(t')} \langle \vec{q}(t) \rangle \right |_{\vec{h}\to \vec{0}}~.
\end{equation}
The stochastic average over paths is needed to compute the response function. In a discrete-time representation for $t_i=i\Delta t$ in $\Delta t\to 0$ limit, the weight functional of a path is given
as proportional to 
$\exp\left[-(4\beta^{-1}\gamma\Delta t)^{-1}\sum_i[ \Delta \vec{W}_i]^2\right]$, where $\Delta \vec{W}_i$ is the Wiener process defined as $\int_{t_i}^{t_{i+1}}ds\vec{\eta}(s)$. From the Langevin equation, $\Delta \vec{W}_i= m(\vec{v}_{i+1}-\vec{v}_i)
+(\mathsf{F}\cdot\vec{r}_i+\mathsf{\Gamma}\cdot\vec{v}_i-\vec{h}_i)\Delta t$
where subscript $i$ denotes a value at $t_i$. It is basically the Onsager-Machlup formalism~\cite{Onsager:1953Fluctuations}.
One can replace $\delta/\delta\vec{h}(t')|_{\vec{h}\to \vec{0}}$ at $t'=t_i$ with the multiplication of $(2\beta^{-1}\gamma)^{-1}\Delta \vec{W}_i/\Delta t$ to $\vec{q}(t)$. Taking the continuous-time limit again, 
\begin{eqnarray}
\lefteqn{\beta^{-1} \mathsf{R}( t, t')=\frac{1}{2\gamma\Delta t}
\langle\vec{q}(t)\Delta \vec{W}(t')\rangle}
\nonumber\\
&=&\frac{1}{2}\langle\vec{q}(t)\vec{v}(t')\rangle+\frac{1}{2\gamma\Delta t}
\langle\vec{q}(t)[\Delta \vec{W}(t')-\gamma\vec{v}(t')\Delta t]\rangle.
\label{response}
\end{eqnarray}
$\mathsf{R}( t, t')=0$ for $t<t'$ because the Wiener process cannot have any influence on $\vec{q}$ at an earlier time, which is known as causality.

The FDR can be examined from $\mathsf{V}(t,t')=\langle\vec{q}(t)\vec{v}(t')\rangle-\beta^{-1}\mathsf{R}(t,t')$ for $t>t'$~\cite{Lee:2017nonequilibrium}. In the following, we use a notation for $4\times 2$ matrices: $[\mathsf{C}]_{\vec{q}\vec{r}}=(C_{i,j})$  and $[\mathsf{C}]_{\vec{q}\vec{v}}=(C_{i,j+2})$ for $1\le i \le 4$ and $j=1,2$. 
For example, $\mathsf{C}_{\vec{q}\vec{r}}(t,t')=\langle \vec{q}(t)\vec{r}(t')\rangle$ and $\mathsf{C}_{\vec{q}\vec{v}}(t,t')=\langle \vec{q}(t)\vec{v}(t')\rangle$. We also let  $\mathsf{V}_{\vec{r}\vec{v}}$ and $\mathsf{V}_{\vec{v}\vec{v}}$ be the upper and the lower block of $\mathsf{V}$, respectively. We can get  
\begin{eqnarray}
\label{eq:violdef}
\lefteqn{\mathsf{V}(t,t')
=\frac{1}{2}\mathsf{C}_{\vec{q}\vec{v}}(t,t')}
\nonumber\\
&&-\frac{1}{2\gamma}\left\langle \vec{q}(t)\left[m\frac{\Delta\vec{v}(t')}{\Delta t} +\mathsf{F}\cdot\vec{r}(t')+b\mathsf{A}\cdot{v}(t')\right]\right\rangle
\end{eqnarray}
where $\Delta \vec{v}(t')=\vec{v}(t'+\Delta t)-\vec{v}(t')$. Note that the term in the square bracket in the above equation is $\Delta \vec{W}(t')/\Delta t-\gamma\vec{v}(t')$, which is the force exerted by the heat bath. 
$\mathsf{V}_{\vec{r}\vec{v}}$ corresponds to the FDR for the position basis~~\cite{Lee:2017nonequilibrium} and $\mathsf{V}_{\vec{v}\vec{v}}$ for the velocity basis~~\cite{Harada:2005equality,Harada:2006energy}. 
In the equal-time limit with $t=t'+\Delta t$, we find
\begin{equation}
\mathsf{V}_{\vec{v}\vec{v}}(t,t)=-\frac{1}{\gamma}\left\langle \frac{\vec{v}(t)+\vec{v}(t')}{2}\left[\frac{\Delta\vec{W}(t')}{\Delta t}-\gamma\vec{v}(t)\right]\right\rangle~,
\end{equation}
for which $\langle\vec{q}(t)\Delta\vec{W}(t')\rangle=\mathsf{0}$ is used. It is the Stratonovich representation for the product $\vec{v}(t)\circ(\vec{\eta}(t)-\gamma\vec{v}(t))$. Then, $\gamma{\rm Tr}\mathsf{V}_{\vec{v}\vec{v}}(t,t)$ is minus the  rate of work done by the reservoir force, which is indeed the rate of heat dissipation.
One can also see $\partial_t\mathsf{V}_{\vec{r}\vec{v}}(t,t')=\mathsf{V}_{\vec{v}\vec{v}}(t,t')$.

In the steady state, $\mathsf{V}(t,t')$ depends only on $\tau=t-t'$ as correlation functions, so written as $\mathsf{V}(\tau)$. 
Using Eqs.~(\ref{correlation_matrix}) and (\ref{eq:violdef}), $\mathsf{V}(\tau)$ can be further simplified as 
\begin{equation}
\begin{split}
\mathsf{V}(\tau)=&\frac{1}{2}\mathsf{C}_{\vec{q}\vec{v}}(\tau) \\
&-\frac{1}{2\gamma}\Big[m[\mathsf{MC(\tau)}]_{\vec{q}\vec{v}}+\mathsf{C}_{\vec{q}\vec{r}}(\tau)\mathsf{F}^{\rm T}-
b\mathsf{C}_{\vec{q}\vec{v}}(\tau)\mathsf{A}\Big]~.
\end{split}
\label{FDR_ss}
\end{equation}
In particular, we find
\begin{equation}
\mathsf{V}(0)=\frac{a}{\beta \gamma \Omega}\left(
\begin{array}{cc} 0 &1\\-1 &0\\ a/\gamma &0\\0&a/\gamma
\end{array}\right)~.
\label{equal_time_FDR}
\end{equation}
$\mathsf{V}(0)=0$ if $a=0$, independent of $b$, which was found in the previous study~\cite{Lee:2017nonequilibrium}. The heat production rate $\langle \dot{Q}\rangle=\gamma{\rm Tr}\mathsf{V}_{\vec{v}\vec{v}}(0)=2a^2\beta^{-1}/(\gamma \Omega)$,
which is consistent with Eq.~(\ref{heat}).

\begin{figure}
\centering
\includegraphics[width=0.45\textwidth]{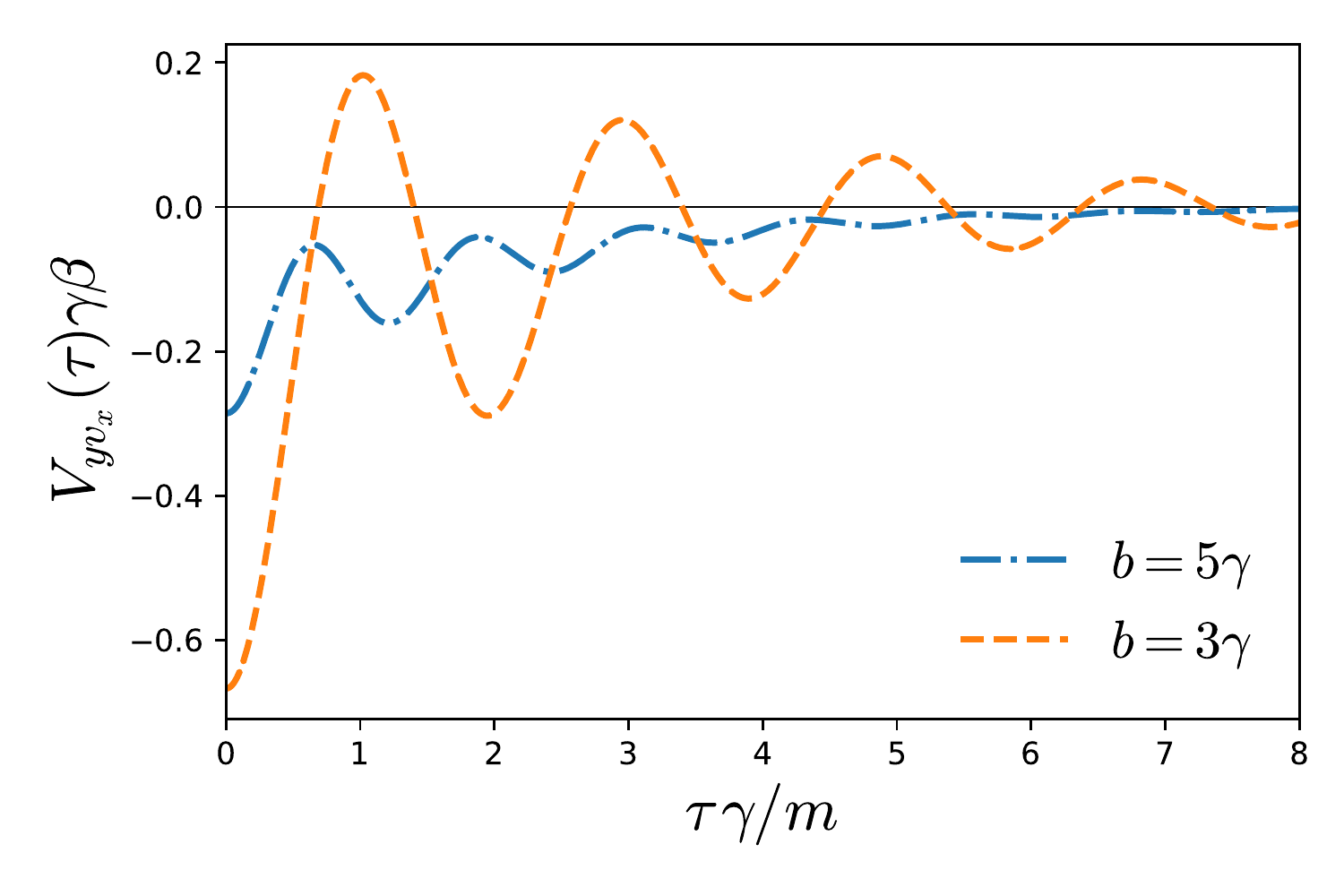}
\caption{\label{fig:violation} The Violation of the FDR in position basis: $V_{yv_x}(\tau)$ for $k = \gamma^2/m$ and $ a = 2 \gamma^2/m $. 
The fluctuation is more amplified for smaller $b$ (smaller $\Omega$).} 
\end{figure}

$\mathsf{V}(\tau)$ in the steady state is given in detail in Appendix~\ref{appendix:violation}. As for the equal-time case in Eq.~(\ref{equal_time_FDR}), $\mathsf{V}(\tau)$ has a multiplicative factor $a/\Omega$. This means the FDR holds only if $a=0$, independent of $b$. It is a quite nontrivial result because $\mathsf{V}(\tau)$ in Eq.~(\ref{FDR_ss}) strongly depends on the correlation functions which differ from those for $b=0$. The DB is violated in this case because $\vec{j}^{\rm irr}\neq 0$; see Eqs.~\eqref{eq:currents} and \eqref{DB}. This result was derived and demonstrated as an example for the exclusiveness of the FDR and the DB, which was derived for a general velocity-dependent force~\cite{Lee:2017nonequilibrium}. For nonzero $a$, both the FDR and the DB are violated. Figure.~\ref{fig:violation} shows  $V_{xv_y}(\tau)$ for given $a$ and increasing $b$. 

\section{Summary} 
\label{summary}
We investigate the combined effect of a non-conservative force generating a torque and a uniform magnetic field on the non-equilibrium motion of a colloid in a harmonic trap. The magnetic field does not work so that the steady state distribution remains Boltzmann in the absence of a non-conservative force. However, combined with the non-conservative force, the magnetic field is found to have a nontrivial influence on the motion out of equilibrium. 

Due to the radial acceleration by the torque, the colloidal motion tends to diverge from the center of the harmonic potential, which might make steady state unreachable. We derive the existence condition for steady state as $k+ab/\gamma-ma^2/\gamma^2>0$. While $a$ (strength of torque) overall tends to depress the possibility of steady state, the magnetic field can enhance it for $ab>0$ (same circulation with the torque) or depress it for $ab<0$ (opposite circulation with the torque).
We find the irreversible current in the velocity space to be composed of the magnetic and (minus) the non-conservative force, which comes up with the circulation current in the position space and the heat production. We find the total entropy change $\Delta S$ to be modified by the unconventional entropy production $\Delta S_{\rm uc}$ originated from a velocity-dependent force, the magnetic field in our study. We find $\Delta S_{uc}$ to measure the dynamical irreversibilty in the circulation current in the position space induced by the magnetic field in combination with the non-conserved force.  We rigorously find two-time correlation functions. In particular, it is noted that the correlation between position and velocity in a perpendicular direction to each other exists as a characteristics of the circulation in the non-equilibrium steady state. We find the combined influence on the violation of the FDR  by the non-conservative force and the magnetic field. 

\begin{acknowledgments} 
This work was supported by the Basic Science Research Program through
the NRF Grant No.~2016R1D1A1A09918020. 
\end{acknowledgments}

\appendix

\section{Two-time Correlation functions in the steady state}
\label{appendix:correl}

From the rotational symmetry for the isotropic case, we can use Eq.~(\ref{eq:isorotsymm}),
$C_{xx}(\tau)=C_{yy}(\tau)$, $C_{xy}(\tau)=-C_{yx}(\tau)$, and also $ \partial_\tau C_{xy}(\tau) = C_{v_xy}(\tau)=- C_{v_yx}(\tau)$. 
For a short notation, we introduce
\begin{equation}
\chi=\frac{1}{\beta\sqrt{R}\Omega}~,~~w+i u=\sqrt{R}e^{i\phi/2},
\end{equation}
where $R$ and $\phi$ are defined in Eqs.~\eqref{eigenvalue} and \eqref{euler}.
Then, from the two basic correlation functions in Eq.~(\ref{eq:twocorr}), we can find remaining correlation functions as follows:
\begin{eqnarray}
\lefteqn{\chi^{-1} C_{xv_x}(\tau)}\\ 
 &=&-(2K+AB-Au) c_1(\tau)-A(w-1)s_1(\tau)\nonumber\\
 && + (2K + AB + Au ) c_2(\tau) +A(w+1) s_2(\tau) \nonumber
\end{eqnarray}
\begin{eqnarray}
 \lefteqn{\chi^{-1} C_{yv_x}(\tau)}\\
 &=& -A( w - 1 ) c_1(\tau)+ (2K +AB -Au )s_1(\tau)\nonumber\\
 &&- A(w+1) c_2(\tau)- ( 2K +AB + Au ) s_2(\tau)\nonumber
 \end{eqnarray}
 \begin{eqnarray}
\lefteqn{\chi^{-1} C_{v_xv_x}(\tau)}\\
&=& -2\left[K( w + 1 ) +  Aw ( B - u )  \right]c_1(\tau)\nonumber\\
&& - \left[ A(w^2 - 1 ) - ( B - u )( 2K + AB -Au ) \right]s_1(\tau)\nonumber\\
&&- 2\left[ K(w-1) + A w ( B + u )\right] c_2(t)\nonumber \\
&&- \left[ -A(w^2 - 1 ) + ( B + u )( 2K + AB + Au ) \right]s_2(t)\nonumber
\end{eqnarray}
\begin{eqnarray}
\lefteqn{\chi^{-1} C_{v_xv_y}(\tau) }\\
&=&\left[(B - u)(2K + AB - Au)-A( w^2 - 1 )\right]c_1(\tau)\nonumber\\
&&+2 \left[K(w+1)+Aw(B - u)\right]s_1(\tau)\nonumber\\
&&-\left[(B + u)(2K + AB + Au) - A( w^2 - 1 )\right]c_2(\tau)\nonumber\\
&&+2\left[K(w - 1) +A(B + u)\right]s_2(\tau)\nonumber
\end{eqnarray}
\bigskip

\section{The violation of FDR in the steady state}
\label{appendix:violation}
We introduce a common factor for the FDR matrix as
\begin{equation}
\kappa=\frac{a}{2\Omega}~.
\end{equation}
Then, the elements of $\mathsf{V}(\tau)$ are found as 
\begin{eqnarray}
\lefteqn{\kappa^{-1}V_{ x v_x }(\tau)}\\
&=&-(2A - B - u)c_1(\tau) 
-(w - 1)s_1(\tau)\nonumber\\
&&  +( 2A -B+u)c_2(\tau) -( w + 1) s_2(\tau) \nonumber
\end{eqnarray}
\begin{eqnarray}
\lefteqn{\kappa^{-1}V_{y v_{x} }(\tau)}\\ 
&=&( 2A- B-u)s_1(\tau) -(w-1)c_1(\tau)\nonumber\\
&& -(2A  - B+u)s_2(\tau)-(w + 1)c_2(\tau)\nonumber
\end{eqnarray}
\begin{eqnarray}
\lefteqn{\kappa^{-1}V_{ v_{x} v_{x} }(\tau)}\\
&=&-( 2 K + A B - A u ) s_1(\tau)+A  (w-1)c_1(\tau) \nonumber\\
&& +( 2 K + A B + Au)s_2(\tau)  + A ( w + 1 )c _2(\tau) \nonumber
\end{eqnarray}
\begin{eqnarray}
\lefteqn{\kappa^{-1}V_{ v_y v_{x} }(\tau)}\\
&=& -(2 K + A B - A u)c_1(\tau)- A(w-1)s_1(\tau) \nonumber\\
&& + (2 K + A B + A u )c_2(\tau) -A(w+1)s_2(\tau) \nonumber
\end{eqnarray}
The FDR is always violated for $a=0$, irrespective of $b$ due to the common multiplicative factor $\kappa$.  

\bibliography{BM-ref}

\begin{thebibliography}{55}%
\makeatletter
\providecommand \@ifxundefined [1]{%
 \@ifx{#1\undefined}
}%
\providecommand \@ifnum [1]{%
 \ifnum #1\expandafter \@firstoftwo
 \else \expandafter \@secondoftwo
 \fi
}%
\providecommand \@ifx [1]{%
 \ifx #1\expandafter \@firstoftwo
 \else \expandafter \@secondoftwo
 \fi
}%
\providecommand \natexlab [1]{#1}%
\providecommand \enquote  [1]{``#1''}%
\providecommand \bibnamefont  [1]{#1}%
\providecommand \bibfnamefont [1]{#1}%
\providecommand \citenamefont [1]{#1}%
\providecommand \href@noop [0]{\@secondoftwo}%
\providecommand \href [0]{\begingroup \@sanitize@url \@href}%
\providecommand \@href[1]{\@@startlink{#1}\@@href}%
\providecommand \@@href[1]{\endgroup#1\@@endlink}%
\providecommand \@sanitize@url [0]{\catcode `\\12\catcode `\$12\catcode
  `\&12\catcode `\#12\catcode `\^12\catcode `\_12\catcode `\%12\relax}%
\providecommand \@@startlink[1]{}%
\providecommand \@@endlink[0]{}%
\providecommand \url  [0]{\begingroup\@sanitize@url \@url }%
\providecommand \@url [1]{\endgroup\@href {#1}{\urlprefix }}%
\providecommand \urlprefix  [0]{URL }%
\providecommand \Eprint [0]{\href }%
\providecommand \doibase [0]{http://dx.doi.org/}%
\providecommand \selectlanguage [0]{\@gobble}%
\providecommand \bibinfo  [0]{\@secondoftwo}%
\providecommand \bibfield  [0]{\@secondoftwo}%
\providecommand \translation [1]{[#1]}%
\providecommand \BibitemOpen [0]{}%
\providecommand \bibitemStop [0]{}%
\providecommand \bibitemNoStop [0]{.\EOS\space}%
\providecommand \EOS [0]{\spacefactor3000\relax}%
\providecommand \BibitemShut  [1]{\csname bibitem#1\endcsname}%
\let\auto@bib@innerbib\@empty
\bibitem [{\citenamefont {Evans}\ \emph {et~al.}(1993)\citenamefont {Evans},
  \citenamefont {Cohen},\ and\ \citenamefont {Morriss}}]{Evans:1993Prob}%
  \BibitemOpen
  \bibfield  {author} {\bibinfo {author} {\bibfnamefont {D.~J.}\ \bibnamefont
  {Evans}}, \bibinfo {author} {\bibfnamefont {E.~G.~D.}\ \bibnamefont {Cohen}},
  \ and\ \bibinfo {author} {\bibfnamefont {G.~P.}\ \bibnamefont {Morriss}},\
  }\href {\doibase 10.1103/PhysRevLett.71.2401} {\bibfield  {journal} {\bibinfo
   {journal} {Phys. Rev. Lett.}\ }\textbf {\bibinfo {volume} {71}},\ \bibinfo
  {pages} {2401} (\bibinfo {year} {1993})}\BibitemShut {NoStop}%
\bibitem [{\citenamefont {Evans}\ and\ \citenamefont
  {Searles}(1994)}]{Evans:1994Equili}%
  \BibitemOpen
  \bibfield  {author} {\bibinfo {author} {\bibfnamefont {D.~J.}\ \bibnamefont
  {Evans}}\ and\ \bibinfo {author} {\bibfnamefont {D.~J.}\ \bibnamefont
  {Searles}},\ }\href {\doibase 10.1103/PhysRevE.50.1645} {\bibfield  {journal}
  {\bibinfo  {journal} {Phys. Rev. E}\ }\textbf {\bibinfo {volume} {50}},\
  \bibinfo {pages} {1645} (\bibinfo {year} {1994})}\BibitemShut {NoStop}%
\bibitem [{\citenamefont {Gallavotti}\ and\ \citenamefont
  {Cohen}(1995{\natexlab{a}})}]{Gallavotti:1995Dynamical}%
  \BibitemOpen
  \bibfield  {author} {\bibinfo {author} {\bibfnamefont {G.}~\bibnamefont
  {Gallavotti}}\ and\ \bibinfo {author} {\bibfnamefont {E.~G.~D.}\ \bibnamefont
  {Cohen}},\ }\href {\doibase 10.1103/PhysRevLett.74.2694} {\bibfield
  {journal} {\bibinfo  {journal} {Phys. Rev. Lett.}\ }\textbf {\bibinfo
  {volume} {74}},\ \bibinfo {pages} {2694} (\bibinfo {year}
  {1995}{\natexlab{a}})}\BibitemShut {NoStop}%
\bibitem [{\citenamefont {Gallavotti}\ and\ \citenamefont
  {Cohen}(1995{\natexlab{b}})}]{Gallavotti:1995dyna}%
  \BibitemOpen
  \bibfield  {author} {\bibinfo {author} {\bibfnamefont {G.}~\bibnamefont
  {Gallavotti}}\ and\ \bibinfo {author} {\bibfnamefont {E.~G.~D.}\ \bibnamefont
  {Cohen}},\ }\href {\doibase 10.1007/BF02179860} {\bibfield  {journal}
  {\bibinfo  {journal} {J. Stat. Phys.}\ }\textbf {\bibinfo {volume} {80}},\
  \bibinfo {pages} {931} (\bibinfo {year} {1995}{\natexlab{b}})}\BibitemShut
  {NoStop}%
\bibitem [{\citenamefont
  {Jarzynski}(1997{\natexlab{a}})}]{Jarzynski:1997Equili}%
  \BibitemOpen
  \bibfield  {author} {\bibinfo {author} {\bibfnamefont {C.}~\bibnamefont
  {Jarzynski}},\ }\href {\doibase 10.1103/PhysRevE.56.5018} {\bibfield
  {journal} {\bibinfo  {journal} {Phys. Rev. E}\ }\textbf {\bibinfo {volume}
  {56}},\ \bibinfo {pages} {5018} (\bibinfo {year}
  {1997}{\natexlab{a}})}\BibitemShut {NoStop}%
\bibitem [{\citenamefont
  {Jarzynski}(1997{\natexlab{b}})}]{Jarzynski:1997Nonequilibrium}%
  \BibitemOpen
  \bibfield  {author} {\bibinfo {author} {\bibfnamefont {C.}~\bibnamefont
  {Jarzynski}},\ }\href {\doibase 10.1103/PhysRevLett.78.2690} {\bibfield
  {journal} {\bibinfo  {journal} {Phys. Rev. Lett.}\ }\textbf {\bibinfo
  {volume} {78}},\ \bibinfo {pages} {2690} (\bibinfo {year}
  {1997}{\natexlab{b}})}\BibitemShut {NoStop}%
\bibitem [{\citenamefont {Kurchan}(1998)}]{Kurchan:1998Fluctuation}%
  \BibitemOpen
  \bibfield  {author} {\bibinfo {author} {\bibfnamefont {J.}~\bibnamefont
  {Kurchan}},\ }\href {http://stacks.iop.org/0305-4470/31/i=16/a=003}
  {\bibfield  {journal} {\bibinfo  {journal} {J. Phys. A}\ }\textbf {\bibinfo
  {volume} {31}},\ \bibinfo {pages} {3719} (\bibinfo {year}
  {1998})}\BibitemShut {NoStop}%
\bibitem [{\citenamefont {Crooks}(1999)}]{Crooks:1999Entropy}%
  \BibitemOpen
  \bibfield  {author} {\bibinfo {author} {\bibfnamefont {G.~E.}\ \bibnamefont
  {Crooks}},\ }\href {\doibase 10.1103/PhysRevE.60.2721} {\bibfield  {journal}
  {\bibinfo  {journal} {Phys. Rev. E}\ }\textbf {\bibinfo {volume} {60}},\
  \bibinfo {pages} {2721} (\bibinfo {year} {1999})}\BibitemShut {NoStop}%
\bibitem [{\citenamefont {Lebowitz}\ and\ \citenamefont
  {Spohn}(1999)}]{Lebowitz:1999Gallavotti}%
  \BibitemOpen
  \bibfield  {author} {\bibinfo {author} {\bibfnamefont {J.~L.}\ \bibnamefont
  {Lebowitz}}\ and\ \bibinfo {author} {\bibfnamefont {H.}~\bibnamefont
  {Spohn}},\ }\href {\doibase 10.1023/A:1004589714161} {\bibfield  {journal}
  {\bibinfo  {journal} {J. Stat. Phys.}\ }\textbf {\bibinfo {volume} {95}},\
  \bibinfo {pages} {333} (\bibinfo {year} {1999})}\BibitemShut {NoStop}%
\bibitem [{\citenamefont {Hatano}\ and\ \citenamefont
  {Sasa}(2001)}]{Hatano:2001Steady}%
  \BibitemOpen
  \bibfield  {author} {\bibinfo {author} {\bibfnamefont {T.}~\bibnamefont
  {Hatano}}\ and\ \bibinfo {author} {\bibfnamefont {S.-i.}\ \bibnamefont
  {Sasa}},\ }\href {\doibase 10.1103/PhysRevLett.86.3463} {\bibfield  {journal}
  {\bibinfo  {journal} {Phys. Rev. Lett.}\ }\textbf {\bibinfo {volume} {86}},\
  \bibinfo {pages} {3463} (\bibinfo {year} {2001})}\BibitemShut {NoStop}%
\bibitem [{\citenamefont {Speck}\ and\ \citenamefont
  {Seifert}(2005)}]{Speck:2005Integral}%
  \BibitemOpen
  \bibfield  {author} {\bibinfo {author} {\bibfnamefont {T.}~\bibnamefont
  {Speck}}\ and\ \bibinfo {author} {\bibfnamefont {U.}~\bibnamefont
  {Seifert}},\ }\href {http://stacks.iop.org/0305-4470/38/i=34/a=L03}
  {\bibfield  {journal} {\bibinfo  {journal} {J. Phys. A}\ }\textbf {\bibinfo
  {volume} {38}},\ \bibinfo {pages} {L581} (\bibinfo {year}
  {2005})}\BibitemShut {NoStop}%
\bibitem [{\citenamefont {Speck}\ and\ \citenamefont
  {Seifert}(2006)}]{Speck:2006restoring}%
  \BibitemOpen
  \bibfield  {author} {\bibinfo {author} {\bibfnamefont {T.}~\bibnamefont
  {Speck}}\ and\ \bibinfo {author} {\bibfnamefont {U.}~\bibnamefont
  {Seifert}},\ }\href@noop {} {\bibfield  {journal} {\bibinfo  {journal} {EPL}\
  }\textbf {\bibinfo {volume} {74}},\ \bibinfo {pages} {391} (\bibinfo {year}
  {2006})}\BibitemShut {NoStop}%
\bibitem [{\citenamefont {Esposito}\ and\ \citenamefont {Van~den
  Broeck}(2010)}]{Esposito:2010Three}%
  \BibitemOpen
  \bibfield  {author} {\bibinfo {author} {\bibfnamefont {M.}~\bibnamefont
  {Esposito}}\ and\ \bibinfo {author} {\bibfnamefont {C.}~\bibnamefont {Van~den
  Broeck}},\ }\href {\doibase 10.1103/PhysRevLett.104.090601} {\bibfield
  {journal} {\bibinfo  {journal} {Phys. Rev. Lett.}\ }\textbf {\bibinfo
  {volume} {104}},\ \bibinfo {pages} {090601} (\bibinfo {year}
  {2010})}\BibitemShut {NoStop}%
\bibitem [{\citenamefont {Wang}\ \emph {et~al.}(2002)\citenamefont {Wang},
  \citenamefont {Sevick}, \citenamefont {Mittag}, \citenamefont {Searles},\
  and\ \citenamefont {Evans}}]{Wang:2002Experi}%
  \BibitemOpen
  \bibfield  {author} {\bibinfo {author} {\bibfnamefont {G.~M.}\ \bibnamefont
  {Wang}}, \bibinfo {author} {\bibfnamefont {E.~M.}\ \bibnamefont {Sevick}},
  \bibinfo {author} {\bibfnamefont {E.}~\bibnamefont {Mittag}}, \bibinfo
  {author} {\bibfnamefont {D.~J.}\ \bibnamefont {Searles}}, \ and\ \bibinfo
  {author} {\bibfnamefont {D.~J.}\ \bibnamefont {Evans}},\ }\href {\doibase
  10.1103/PhysRevLett.89.050601} {\bibfield  {journal} {\bibinfo  {journal}
  {Phys. Rev. Lett.}\ }\textbf {\bibinfo {volume} {89}},\ \bibinfo {pages}
  {050601} (\bibinfo {year} {2002})}\BibitemShut {NoStop}%
\bibitem [{\citenamefont {Hummer}\ and\ \citenamefont
  {Szabo}(2001)}]{Hummer:2001Free}%
  \BibitemOpen
  \bibfield  {author} {\bibinfo {author} {\bibfnamefont {G.}~\bibnamefont
  {Hummer}}\ and\ \bibinfo {author} {\bibfnamefont {A.}~\bibnamefont {Szabo}},\
  }\href {\doibase 10.1073/pnas.071034098} {\bibfield  {journal} {\bibinfo
  {journal} {Proc. Natl. Acad. Sci.}\ }\textbf {\bibinfo {volume} {98}},\
  \bibinfo {pages} {3658} (\bibinfo {year} {2001})}\BibitemShut {NoStop}%
\bibitem [{\citenamefont {Liphardt}\ \emph {et~al.}(2002)\citenamefont
  {Liphardt}, \citenamefont {Dumont}, \citenamefont {Smith}, \citenamefont
  {Tinoco},\ and\ \citenamefont {Bustamante}}]{Liphardt:2002Equili}%
  \BibitemOpen
  \bibfield  {author} {\bibinfo {author} {\bibfnamefont {J.}~\bibnamefont
  {Liphardt}}, \bibinfo {author} {\bibfnamefont {S.}~\bibnamefont {Dumont}},
  \bibinfo {author} {\bibfnamefont {S.~B.}\ \bibnamefont {Smith}}, \bibinfo
  {author} {\bibfnamefont {I.}~\bibnamefont {Tinoco}}, \ and\ \bibinfo {author}
  {\bibfnamefont {C.}~\bibnamefont {Bustamante}},\ }\href {\doibase
  10.1126/science.1071152} {\bibfield  {journal} {\bibinfo  {journal}
  {Science}\ }\textbf {\bibinfo {volume} {296}},\ \bibinfo {pages} {1832}
  (\bibinfo {year} {2002})}\BibitemShut {NoStop}%
\bibitem [{\citenamefont {Collin}\ \emph {et~al.}(2005)\citenamefont {Collin},
  \citenamefont {Ritort}, \citenamefont {Jarzynski}, \citenamefont {Smith},
  \citenamefont {Tinoco~Jr},\ and\ \citenamefont
  {Bustamante}}]{Collin:2005Verifi}%
  \BibitemOpen
  \bibfield  {author} {\bibinfo {author} {\bibfnamefont {D.}~\bibnamefont
  {Collin}}, \bibinfo {author} {\bibfnamefont {F.}~\bibnamefont {Ritort}},
  \bibinfo {author} {\bibfnamefont {C.}~\bibnamefont {Jarzynski}}, \bibinfo
  {author} {\bibfnamefont {S.~B.}\ \bibnamefont {Smith}}, \bibinfo {author}
  {\bibfnamefont {I.}~\bibnamefont {Tinoco~Jr}}, \ and\ \bibinfo {author}
  {\bibfnamefont {C.}~\bibnamefont {Bustamante}},\ }\href
  {https://doi.org/10.1038/nature04061} {\bibfield  {journal} {\bibinfo
  {journal} {Nature}\ }\textbf {\bibinfo {volume} {437}},\ \bibinfo {pages}
  {231} (\bibinfo {year} {2005})}\BibitemShut {NoStop}%
\bibitem [{\citenamefont {Trepagnier}\ \emph {et~al.}(2004)\citenamefont
  {Trepagnier}, \citenamefont {Jarzynski}, \citenamefont {Ritort},
  \citenamefont {Crooks}, \citenamefont {Bustamante},\ and\ \citenamefont
  {Liphardt}}]{Trepagnier:2004Experi}%
  \BibitemOpen
  \bibfield  {author} {\bibinfo {author} {\bibfnamefont {E.~H.}\ \bibnamefont
  {Trepagnier}}, \bibinfo {author} {\bibfnamefont {C.}~\bibnamefont
  {Jarzynski}}, \bibinfo {author} {\bibfnamefont {F.}~\bibnamefont {Ritort}},
  \bibinfo {author} {\bibfnamefont {G.~E.}\ \bibnamefont {Crooks}}, \bibinfo
  {author} {\bibfnamefont {C.~J.}\ \bibnamefont {Bustamante}}, \ and\ \bibinfo
  {author} {\bibfnamefont {J.}~\bibnamefont {Liphardt}},\ }\href {\doibase
  10.1073/pnas.0406405101} {\bibfield  {journal} {\bibinfo  {journal} {Proc.
  Natl. Acad. Sci.}\ }\textbf {\bibinfo {volume} {101}},\ \bibinfo {pages}
  {15038} (\bibinfo {year} {2004})}\BibitemShut {NoStop}%
\bibitem [{\citenamefont {Garnier}\ and\ \citenamefont
  {Ciliberto}(2005)}]{Garnier:2005Nonequili}%
  \BibitemOpen
  \bibfield  {author} {\bibinfo {author} {\bibfnamefont {N.}~\bibnamefont
  {Garnier}}\ and\ \bibinfo {author} {\bibfnamefont {S.}~\bibnamefont
  {Ciliberto}},\ }\href {\doibase 10.1103/PhysRevE.71.060101} {\bibfield
  {journal} {\bibinfo  {journal} {Phys. Rev. E}\ }\textbf {\bibinfo {volume}
  {71}},\ \bibinfo {pages} {060101} (\bibinfo {year} {2005})}\BibitemShut
  {NoStop}%
\bibitem [{\citenamefont {Douarche}\ \emph {et~al.}(2006)\citenamefont
  {Douarche}, \citenamefont {Joubaud}, \citenamefont {Garnier}, \citenamefont
  {Petrosyan},\ and\ \citenamefont {Ciliberto}}]{Douarche:2006Work}%
  \BibitemOpen
  \bibfield  {author} {\bibinfo {author} {\bibfnamefont {F.}~\bibnamefont
  {Douarche}}, \bibinfo {author} {\bibfnamefont {S.}~\bibnamefont {Joubaud}},
  \bibinfo {author} {\bibfnamefont {N.~B.}\ \bibnamefont {Garnier}}, \bibinfo
  {author} {\bibfnamefont {A.}~\bibnamefont {Petrosyan}}, \ and\ \bibinfo
  {author} {\bibfnamefont {S.}~\bibnamefont {Ciliberto}},\ }\href {\doibase
  10.1103/PhysRevLett.97.140603} {\bibfield  {journal} {\bibinfo  {journal}
  {Phys. Rev. Lett.}\ }\textbf {\bibinfo {volume} {97}},\ \bibinfo {pages}
  {140603} (\bibinfo {year} {2006})}\BibitemShut {NoStop}%
\bibitem [{\citenamefont {Joubaud}\ \emph {et~al.}(2008)\citenamefont
  {Joubaud}, \citenamefont {Garnier},\ and\ \citenamefont
  {Ciliberto}}]{Joubaud:2008Fluc}%
  \BibitemOpen
  \bibfield  {author} {\bibinfo {author} {\bibfnamefont {S.}~\bibnamefont
  {Joubaud}}, \bibinfo {author} {\bibfnamefont {N.~B.}\ \bibnamefont
  {Garnier}}, \ and\ \bibinfo {author} {\bibfnamefont {S.}~\bibnamefont
  {Ciliberto}},\ }\href@noop {} {\bibfield  {journal} {\bibinfo  {journal}
  {EPL}\ }\textbf {\bibinfo {volume} {82}},\ \bibinfo {pages} {30007} (\bibinfo
  {year} {2008})}\BibitemShut {NoStop}%
\bibitem [{\citenamefont {Hayashi}\ \emph {et~al.}(2010)\citenamefont
  {Hayashi}, \citenamefont {Ueno}, \citenamefont {Iino},\ and\ \citenamefont
  {Noji}}]{Hayashi:2010Fluc}%
  \BibitemOpen
  \bibfield  {author} {\bibinfo {author} {\bibfnamefont {K.}~\bibnamefont
  {Hayashi}}, \bibinfo {author} {\bibfnamefont {H.}~\bibnamefont {Ueno}},
  \bibinfo {author} {\bibfnamefont {R.}~\bibnamefont {Iino}}, \ and\ \bibinfo
  {author} {\bibfnamefont {H.}~\bibnamefont {Noji}},\ }\href {\doibase
  10.1103/PhysRevLett.104.218103} {\bibfield  {journal} {\bibinfo  {journal}
  {Phys. Rev. Lett.}\ }\textbf {\bibinfo {volume} {104}},\ \bibinfo {pages}
  {218103} (\bibinfo {year} {2010})}\BibitemShut {NoStop}%
\bibitem [{\citenamefont {Lee}\ \emph {et~al.}(2015)\citenamefont {Lee},
  \citenamefont {Kwon},\ and\ \citenamefont {Pak}}]{Lee:2015Nonequili}%
  \BibitemOpen
  \bibfield  {author} {\bibinfo {author} {\bibfnamefont {D.~Y.}\ \bibnamefont
  {Lee}}, \bibinfo {author} {\bibfnamefont {C.}~\bibnamefont {Kwon}}, \ and\
  \bibinfo {author} {\bibfnamefont {H.~K.}\ \bibnamefont {Pak}},\ }\href
  {\doibase 10.1103/PhysRevLett.114.060603} {\bibfield  {journal} {\bibinfo
  {journal} {Phys. Rev. Lett.}\ }\textbf {\bibinfo {volume} {114}},\ \bibinfo
  {pages} {060603} (\bibinfo {year} {2015})}\BibitemShut {NoStop}%
\bibitem [{\citenamefont {Taylor}(1961)}]{Taylor:1961diffusion}%
  \BibitemOpen
  \bibfield  {author} {\bibinfo {author} {\bibfnamefont {J.}~\bibnamefont
  {Taylor}},\ }\href@noop {} {\bibfield  {journal} {\bibinfo  {journal} {Phys.
  Fluids}\ }\textbf {\bibinfo {volume} {4}},\ \bibinfo {pages} {1142} (\bibinfo
  {year} {1961})}\BibitemShut {NoStop}%
\bibitem [{\citenamefont {Kurşunoǧlu}(1963)}]{Kursunoglu:1963brownian}%
  \BibitemOpen
  \bibfield  {author} {\bibinfo {author} {\bibfnamefont {B.}~\bibnamefont
  {Kurşunoǧlu}},\ }\href@noop {} {\bibfield  {journal} {\bibinfo  {journal}
  {Phys. Rev.}\ }\textbf {\bibinfo {volume} {132}},\ \bibinfo {pages} {21}
  (\bibinfo {year} {1963})}\BibitemShut {NoStop}%
\bibitem [{\citenamefont {Xiang}(1993)}]{Xiang:1993stochastic}%
  \BibitemOpen
  \bibfield  {author} {\bibinfo {author} {\bibfnamefont {N.}~\bibnamefont
  {Xiang}},\ }\href@noop {} {\bibfield  {journal} {\bibinfo  {journal} {Phys.
  Rev. E}\ }\textbf {\bibinfo {volume} {48}},\ \bibinfo {pages} {1590}
  (\bibinfo {year} {1993})}\BibitemShut {NoStop}%
\bibitem [{\citenamefont {Balescu}(1997)}]{Balescu:1997matter}%
  \BibitemOpen
  \bibfield  {author} {\bibinfo {author} {\bibfnamefont {R.}~\bibnamefont
  {Balescu}},\ }\href@noop {} {\emph {\bibinfo {title} {Statistical dynamics:
  matter out of equilibrium}}}\ (\bibinfo  {publisher} {Imperial Coll.},\
  \bibinfo {year} {1997})\BibitemShut {NoStop}%
\bibitem [{\citenamefont {Czopnik}\ and\ \citenamefont
  {Garbaczewski}(2001)}]{Czopnik:2001brownian}%
  \BibitemOpen
  \bibfield  {author} {\bibinfo {author} {\bibfnamefont {R.}~\bibnamefont
  {Czopnik}}\ and\ \bibinfo {author} {\bibfnamefont {P.}~\bibnamefont
  {Garbaczewski}},\ }\href@noop {} {\bibfield  {journal} {\bibinfo  {journal}
  {Phys. Rev. E}\ }\textbf {\bibinfo {volume} {63}},\ \bibinfo {pages} {021105}
  (\bibinfo {year} {2001})}\BibitemShut {NoStop}%
\bibitem [{\citenamefont {Jiménez-Aquino}\ and\ \citenamefont
  {Romero-Bastida}(2007)}]{Jimenez:2007fokker}%
  \BibitemOpen
  \bibfield  {author} {\bibinfo {author} {\bibfnamefont {J.}~\bibnamefont
  {Jiménez-Aquino}}\ and\ \bibinfo {author} {\bibfnamefont {M.}~\bibnamefont
  {Romero-Bastida}},\ }\href@noop {} {\bibfield  {journal} {\bibinfo  {journal}
  {Phys. Rev. E}\ }\textbf {\bibinfo {volume} {76}},\ \bibinfo {pages} {021106}
  (\bibinfo {year} {2007})}\BibitemShut {NoStop}%
\bibitem [{\citenamefont {Jayannavar}\ and\ \citenamefont
  {Sahoo}(2007)}]{Jayannavar:2007charged}%
  \BibitemOpen
  \bibfield  {author} {\bibinfo {author} {\bibfnamefont {A.}~\bibnamefont
  {Jayannavar}}\ and\ \bibinfo {author} {\bibfnamefont {M.}~\bibnamefont
  {Sahoo}},\ }\href@noop {} {\bibfield  {journal} {\bibinfo  {journal} {Phys.
  Rev. E}\ }\textbf {\bibinfo {volume} {75}},\ \bibinfo {pages} {032102}
  (\bibinfo {year} {2007})}\BibitemShut {NoStop}%
\bibitem [{\citenamefont {Jiménez-Aquino}\ \emph {et~al.}(2010)\citenamefont
  {Jiménez-Aquino}, \citenamefont {Uribe},\ and\ \citenamefont
  {Velasco}}]{Jimenez:2010work}%
  \BibitemOpen
  \bibfield  {author} {\bibinfo {author} {\bibfnamefont {J.}~\bibnamefont
  {Jiménez-Aquino}}, \bibinfo {author} {\bibfnamefont {F.}~\bibnamefont
  {Uribe}}, \ and\ \bibinfo {author} {\bibfnamefont {R.}~\bibnamefont
  {Velasco}},\ }\href@noop {} {\bibfield  {journal} {\bibinfo  {journal} {J.
  Phys. A}\ }\textbf {\bibinfo {volume} {43}},\ \bibinfo {pages} {255001}
  (\bibinfo {year} {2010})}\BibitemShut {NoStop}%
\bibitem [{\citenamefont {Jiménez-Aquino}\ and\ \citenamefont
  {Romero-Bastida}(2013)}]{Jimenez:2013brownian}%
  \BibitemOpen
  \bibfield  {author} {\bibinfo {author} {\bibfnamefont {J.}~\bibnamefont
  {Jiménez-Aquino}}\ and\ \bibinfo {author} {\bibfnamefont {M.}~\bibnamefont
  {Romero-Bastida}},\ }\href@noop {} {\bibfield  {journal} {\bibinfo  {journal}
  {Phys. Rev. E}\ }\textbf {\bibinfo {volume} {88}},\ \bibinfo {pages} {022151}
  (\bibinfo {year} {2013})}\BibitemShut {NoStop}%
\bibitem [{\citenamefont {Kwon}\ \emph {et~al.}(2016)\citenamefont {Kwon},
  \citenamefont {Yeo}, \citenamefont {Lee},\ and\ \citenamefont
  {Park}}]{Kwon:2016unconventional}%
  \BibitemOpen
  \bibfield  {author} {\bibinfo {author} {\bibfnamefont {C.}~\bibnamefont
  {Kwon}}, \bibinfo {author} {\bibfnamefont {J.}~\bibnamefont {Yeo}}, \bibinfo
  {author} {\bibfnamefont {H.~K.}\ \bibnamefont {Lee}}, \ and\ \bibinfo
  {author} {\bibfnamefont {H.}~\bibnamefont {Park}},\ }\href@noop {} {\bibfield
   {journal} {\bibinfo  {journal} {J. Korean Phys. Soc.}\ }\textbf {\bibinfo
  {volume} {68}},\ \bibinfo {pages} {633} (\bibinfo {year} {2016})}\BibitemShut
  {NoStop}%
\bibitem [{\citenamefont {Chun}\ \emph {et~al.}(2018)\citenamefont {Chun},
  \citenamefont {Durang},\ and\ \citenamefont {Noh}}]{Chun:2018emergence}%
  \BibitemOpen
  \bibfield  {author} {\bibinfo {author} {\bibfnamefont {H.-M.}\ \bibnamefont
  {Chun}}, \bibinfo {author} {\bibfnamefont {X.}~\bibnamefont {Durang}}, \ and\
  \bibinfo {author} {\bibfnamefont {J.~D.}\ \bibnamefont {Noh}},\ }\href@noop
  {} {\bibfield  {journal} {\bibinfo  {journal} {Phys. Rev. E}\ }\textbf
  {\bibinfo {volume} {97}},\ \bibinfo {pages} {032117} (\bibinfo {year}
  {2018})}\BibitemShut {NoStop}%
\bibitem [{\citenamefont {Kwon}\ \emph {et~al.}(2011)\citenamefont {Kwon},
  \citenamefont {Noh},\ and\ \citenamefont {Park}}]{Kwon:2011nonequilibrium}%
  \BibitemOpen
  \bibfield  {author} {\bibinfo {author} {\bibfnamefont {C.}~\bibnamefont
  {Kwon}}, \bibinfo {author} {\bibfnamefont {J.~D.}\ \bibnamefont {Noh}}, \
  and\ \bibinfo {author} {\bibfnamefont {H.}~\bibnamefont {Park}},\ }\href@noop
  {} {\bibfield  {journal} {\bibinfo  {journal} {Phys. Rev. E}\ }\textbf
  {\bibinfo {volume} {83}},\ \bibinfo {pages} {061145} (\bibinfo {year}
  {2011})}\BibitemShut {NoStop}%
\bibitem [{\citenamefont {Noh}\ \emph {et~al.}(2013)\citenamefont {Noh},
  \citenamefont {Kwon},\ and\ \citenamefont {Park}}]{Noh:2013Multiple}%
  \BibitemOpen
  \bibfield  {author} {\bibinfo {author} {\bibfnamefont {J.~D.}\ \bibnamefont
  {Noh}}, \bibinfo {author} {\bibfnamefont {C.}~\bibnamefont {Kwon}}, \ and\
  \bibinfo {author} {\bibfnamefont {H.}~\bibnamefont {Park}},\ }\href {\doibase
  10.1103/PhysRevLett.111.130601} {\bibfield  {journal} {\bibinfo  {journal}
  {Phys. Rev. Lett.}\ }\textbf {\bibinfo {volume} {111}},\ \bibinfo {pages}
  {130601} (\bibinfo {year} {2013})}\BibitemShut {NoStop}%
\bibitem [{\citenamefont {Kwon}\ \emph {et~al.}(2005)\citenamefont {Kwon},
  \citenamefont {Ao},\ and\ \citenamefont {Thouless}}]{Kwon:2005structure}%
  \BibitemOpen
  \bibfield  {author} {\bibinfo {author} {\bibfnamefont {C.}~\bibnamefont
  {Kwon}}, \bibinfo {author} {\bibfnamefont {P.}~\bibnamefont {Ao}}, \ and\
  \bibinfo {author} {\bibfnamefont {D.~J.}\ \bibnamefont {Thouless}},\
  }\href@noop {} {\bibfield  {journal} {\bibinfo  {journal} {Proc. Natl. Acad.
  Sci. U. S. A.}\ }\textbf {\bibinfo {volume} {102}},\ \bibinfo {pages} {13029}
  (\bibinfo {year} {2005})}\BibitemShut {NoStop}%
\bibitem [{\citenamefont {Filliger}\ and\ \citenamefont
  {Reimann}(2007)}]{Filliger:2007Brown}%
  \BibitemOpen
  \bibfield  {author} {\bibinfo {author} {\bibfnamefont {R.}~\bibnamefont
  {Filliger}}\ and\ \bibinfo {author} {\bibfnamefont {P.}~\bibnamefont
  {Reimann}},\ }\href {\doibase 10.1103/PhysRevLett.99.230602} {\bibfield
  {journal} {\bibinfo  {journal} {Phys. Rev. Lett.}\ }\textbf {\bibinfo
  {volume} {99}},\ \bibinfo {pages} {230602} (\bibinfo {year}
  {2007})}\BibitemShut {NoStop}%
\bibitem [{\citenamefont {Park}\ \emph {et~al.}(2016)\citenamefont {Park},
  \citenamefont {Chun},\ and\ \citenamefont {Noh}}]{Park:2016efficiency}%
  \BibitemOpen
  \bibfield  {author} {\bibinfo {author} {\bibfnamefont {J.-M.}\ \bibnamefont
  {Park}}, \bibinfo {author} {\bibfnamefont {H.-M.}\ \bibnamefont {Chun}}, \
  and\ \bibinfo {author} {\bibfnamefont {J.~D.}\ \bibnamefont {Noh}},\ }\href
  {\doibase 10.1103/PhysRevE.94.012127} {\bibfield  {journal} {\bibinfo
  {journal} {Phys. Rev. E}\ }\textbf {\bibinfo {volume} {94}},\ \bibinfo
  {pages} {012127} (\bibinfo {year} {2016})}\BibitemShut {NoStop}%
\bibitem [{\citenamefont {Ciliberto}\ \emph {et~al.}(2013)\citenamefont
  {Ciliberto}, \citenamefont {Imparato}, \citenamefont {Naert},\ and\
  \citenamefont {Tanase}}]{Ciliberto:2013Heat}%
  \BibitemOpen
  \bibfield  {author} {\bibinfo {author} {\bibfnamefont {S.}~\bibnamefont
  {Ciliberto}}, \bibinfo {author} {\bibfnamefont {A.}~\bibnamefont {Imparato}},
  \bibinfo {author} {\bibfnamefont {A.}~\bibnamefont {Naert}}, \ and\ \bibinfo
  {author} {\bibfnamefont {M.}~\bibnamefont {Tanase}},\ }\href {\doibase
  10.1103/PhysRevLett.110.180601} {\bibfield  {journal} {\bibinfo  {journal}
  {Phys. Rev. Lett.}\ }\textbf {\bibinfo {volume} {110}},\ \bibinfo {pages}
  {180601} (\bibinfo {year} {2013})}\BibitemShut {NoStop}%
\bibitem [{\citenamefont {Risken}(1996)}]{Risken:1996fokker}%
  \BibitemOpen
  \bibfield  {author} {\bibinfo {author} {\bibfnamefont {H.}~\bibnamefont
  {Risken}},\ }\href@noop {} {\emph {\bibinfo {title} {The Fokker-Planck
  Equation}}}\ (\bibinfo  {publisher} {Springer},\ \bibinfo {year}
  {1996})\BibitemShut {NoStop}%
\bibitem [{\citenamefont {Lee}\ \emph {et~al.}(2013)\citenamefont {Lee},
  \citenamefont {Kwon},\ and\ \citenamefont {Park}}]{Lee:2013Fluc}%
  \BibitemOpen
  \bibfield  {author} {\bibinfo {author} {\bibfnamefont {H.~K.}\ \bibnamefont
  {Lee}}, \bibinfo {author} {\bibfnamefont {C.}~\bibnamefont {Kwon}}, \ and\
  \bibinfo {author} {\bibfnamefont {H.}~\bibnamefont {Park}},\ }\href {\doibase
  10.1103/PhysRevLett.110.050602} {\bibfield  {journal} {\bibinfo  {journal}
  {Phys. Rev. Lett.}\ }\textbf {\bibinfo {volume} {110}},\ \bibinfo {pages}
  {050602} (\bibinfo {year} {2013})}\BibitemShut {NoStop}%
\bibitem [{\citenamefont {Marchetti}\ \emph {et~al.}(2013)\citenamefont
  {Marchetti}, \citenamefont {Joanny}, \citenamefont {Ramaswamy}, \citenamefont
  {Liverpool}, \citenamefont {Prost}, \citenamefont {Rao},\ and\ \citenamefont
  {Simha}}]{Marchetti:2013Hydro}%
  \BibitemOpen
  \bibfield  {author} {\bibinfo {author} {\bibfnamefont {M.~C.}\ \bibnamefont
  {Marchetti}}, \bibinfo {author} {\bibfnamefont {J.~F.}\ \bibnamefont
  {Joanny}}, \bibinfo {author} {\bibfnamefont {S.}~\bibnamefont {Ramaswamy}},
  \bibinfo {author} {\bibfnamefont {T.~B.}\ \bibnamefont {Liverpool}}, \bibinfo
  {author} {\bibfnamefont {J.}~\bibnamefont {Prost}}, \bibinfo {author}
  {\bibfnamefont {M.}~\bibnamefont {Rao}}, \ and\ \bibinfo {author}
  {\bibfnamefont {R.~A.}\ \bibnamefont {Simha}},\ }\href@noop {} {\bibfield
  {journal} {\bibinfo  {journal} {Rev. Mod. Phys.}\ }\textbf {\bibinfo {volume}
  {85}},\ \bibinfo {pages} {1143} (\bibinfo {year} {2013})}\BibitemShut
  {NoStop}%
\bibitem [{\citenamefont {Kim}\ and\ \citenamefont
  {Qian}(2004)}]{Kim:2004Entropy}%
  \BibitemOpen
  \bibfield  {author} {\bibinfo {author} {\bibfnamefont {K.~H.}\ \bibnamefont
  {Kim}}\ and\ \bibinfo {author} {\bibfnamefont {H.}~\bibnamefont {Qian}},\
  }\href@noop {} {\bibfield  {journal} {\bibinfo  {journal} {Phys. Rev. Lett.}\
  }\textbf {\bibinfo {volume} {93}},\ \bibinfo {pages} {120602} (\bibinfo
  {year} {2004})}\BibitemShut {NoStop}%
\bibitem [{\citenamefont {Kim}\ and\ \citenamefont
  {Qian}(2007)}]{Kim:2007Fluc}%
  \BibitemOpen
  \bibfield  {author} {\bibinfo {author} {\bibfnamefont {K.~H.}\ \bibnamefont
  {Kim}}\ and\ \bibinfo {author} {\bibfnamefont {H.}~\bibnamefont {Qian}},\
  }\href@noop {} {\bibfield  {journal} {\bibinfo  {journal} {Phys. Rev. E}\
  }\textbf {\bibinfo {volume} {75}},\ \bibinfo {pages} {022102} (\bibinfo
  {year} {2007})}\BibitemShut {NoStop}%
\bibitem [{\citenamefont {Schweitzer}(2007)}]{Schweitzer:2007brownian}%
  \BibitemOpen
  \bibfield  {author} {\bibinfo {author} {\bibfnamefont {F.}~\bibnamefont
  {Schweitzer}},\ }\href@noop {} {\emph {\bibinfo {title} {Brownian agents and
  active particles: collective dynamics in the natural and social sciences}}}\
  (\bibinfo  {publisher} {Springer},\ \bibinfo {year} {2007})\BibitemShut
  {NoStop}%
\bibitem [{\citenamefont {Romanczuk}\ \emph {et~al.}(2012)\citenamefont
  {Romanczuk}, \citenamefont {B{\"a}r}, \citenamefont {Ebeling}, \citenamefont
  {Lindner},\ and\ \citenamefont {Schimansky-Geier}}]{Romanczuk:2012Active}%
  \BibitemOpen
  \bibfield  {author} {\bibinfo {author} {\bibfnamefont {P.}~\bibnamefont
  {Romanczuk}}, \bibinfo {author} {\bibfnamefont {M.}~\bibnamefont {B{\"a}r}},
  \bibinfo {author} {\bibfnamefont {W.}~\bibnamefont {Ebeling}}, \bibinfo
  {author} {\bibfnamefont {B.}~\bibnamefont {Lindner}}, \ and\ \bibinfo
  {author} {\bibfnamefont {L.}~\bibnamefont {Schimansky-Geier}},\ }\href@noop
  {} {\bibfield  {journal} {\bibinfo  {journal} {Eur. Phys. J. Spec. Top.}\
  }\textbf {\bibinfo {volume} {202}},\ \bibinfo {pages} {1} (\bibinfo {year}
  {2012})}\BibitemShut {NoStop}%
\bibitem [{\citenamefont {Ganguly}\ and\ \citenamefont
  {Chaudhuri}(2013)}]{Ganguly:2013Stochastic}%
  \BibitemOpen
  \bibfield  {author} {\bibinfo {author} {\bibfnamefont {C.}~\bibnamefont
  {Ganguly}}\ and\ \bibinfo {author} {\bibfnamefont {D.}~\bibnamefont
  {Chaudhuri}},\ }\href@noop {} {\bibfield  {journal} {\bibinfo  {journal}
  {Phys. Rev. E}\ }\textbf {\bibinfo {volume} {88}},\ \bibinfo {pages} {032102}
  (\bibinfo {year} {2013})}\BibitemShut {NoStop}%
\bibitem [{\citenamefont {Chaudhuri}(2014)}]{Chaudhuri:2014Active}%
  \BibitemOpen
  \bibfield  {author} {\bibinfo {author} {\bibfnamefont {D.}~\bibnamefont
  {Chaudhuri}},\ }\href@noop {} {\bibfield  {journal} {\bibinfo  {journal}
  {Phys. Rev. E}\ }\textbf {\bibinfo {volume} {90}},\ \bibinfo {pages} {022131}
  (\bibinfo {year} {2014})}\BibitemShut {NoStop}%
\bibitem [{\citenamefont {{Courty, J.-M.}}\ \emph {et~al.}(2001)\citenamefont
  {{Courty, J.-M.}}, \citenamefont {{Heidmann, A.}},\ and\ \citenamefont
  {{Pinard, M.}}}]{Courty:2001Quantum}%
  \BibitemOpen
  \bibfield  {author} {\bibinfo {author} {\bibnamefont {{Courty, J.-M.}}},
  \bibinfo {author} {\bibnamefont {{Heidmann, A.}}}, \ and\ \bibinfo {author}
  {\bibnamefont {{Pinard, M.}}},\ }\href@noop {} {\bibfield  {journal}
  {\bibinfo  {journal} {Eur. Phys. J. D}\ }\textbf {\bibinfo {volume} {17}},\
  \bibinfo {pages} {399} (\bibinfo {year} {2001})}\BibitemShut {NoStop}%
\bibitem [{\citenamefont {Jourdan}\ \emph {et~al.}(2007)\citenamefont
  {Jourdan}, \citenamefont {Torricelli}, \citenamefont {Chevrier},\ and\
  \citenamefont {Comin}}]{G:2007Tuning}%
  \BibitemOpen
  \bibfield  {author} {\bibinfo {author} {\bibfnamefont {G.}~\bibnamefont
  {Jourdan}}, \bibinfo {author} {\bibfnamefont {G.}~\bibnamefont {Torricelli}},
  \bibinfo {author} {\bibfnamefont {J.}~\bibnamefont {Chevrier}}, \ and\
  \bibinfo {author} {\bibfnamefont {F.}~\bibnamefont {Comin}},\ }\href@noop {}
  {\bibfield  {journal} {\bibinfo  {journal} {Nanotechnology}\ }\textbf
  {\bibinfo {volume} {18}},\ \bibinfo {pages} {475502} (\bibinfo {year}
  {2007})}\BibitemShut {NoStop}%
\bibitem [{\citenamefont {Harada}\ and\ \citenamefont
  {Sasa}(2005)}]{Harada:2005equality}%
  \BibitemOpen
  \bibfield  {author} {\bibinfo {author} {\bibfnamefont {T.}~\bibnamefont
  {Harada}}\ and\ \bibinfo {author} {\bibfnamefont {S.-i.}\ \bibnamefont
  {Sasa}},\ }\href@noop {} {\bibfield  {journal} {\bibinfo  {journal} {Phys.
  Rev. Lett.}\ }\textbf {\bibinfo {volume} {95}},\ \bibinfo {pages} {130602}
  (\bibinfo {year} {2005})}\BibitemShut {NoStop}%
\bibitem [{\citenamefont {Harada}\ and\ \citenamefont
  {Sasa}(2006)}]{Harada:2006energy}%
  \BibitemOpen
  \bibfield  {author} {\bibinfo {author} {\bibfnamefont {T.}~\bibnamefont
  {Harada}}\ and\ \bibinfo {author} {\bibfnamefont {S.-i.}\ \bibnamefont
  {Sasa}},\ }\href@noop {} {\bibfield  {journal} {\bibinfo  {journal} {Phys.
  Rev. E}\ }\textbf {\bibinfo {volume} {73}},\ \bibinfo {pages} {026131}
  (\bibinfo {year} {2006})}\BibitemShut {NoStop}%
\bibitem [{\citenamefont {Lee}\ \emph {et~al.}(2017)\citenamefont {Lee},
  \citenamefont {Lahiri},\ and\ \citenamefont {Park}}]{Lee:2017nonequilibrium}%
  \BibitemOpen
  \bibfield  {author} {\bibinfo {author} {\bibfnamefont {H.~K.}\ \bibnamefont
  {Lee}}, \bibinfo {author} {\bibfnamefont {S.}~\bibnamefont {Lahiri}}, \ and\
  \bibinfo {author} {\bibfnamefont {H.}~\bibnamefont {Park}},\ }\href@noop {}
  {\bibfield  {journal} {\bibinfo  {journal} {Phys. Rev. E}\ }\textbf {\bibinfo
  {volume} {96}},\ \bibinfo {pages} {022134} (\bibinfo {year}
  {2017})}\BibitemShut {NoStop}%
\bibitem [{\citenamefont {Onsager}\ and\ \citenamefont
  {Machlup}(1953)}]{Onsager:1953Fluctuations}%
  \BibitemOpen
  \bibfield  {author} {\bibinfo {author} {\bibfnamefont {L.}~\bibnamefont
  {Onsager}}\ and\ \bibinfo {author} {\bibfnamefont {S.}~\bibnamefont
  {Machlup}},\ }\href@noop {} {\bibfield  {journal} {\bibinfo  {journal} {Phys.
  Rev.}\ }\textbf {\bibinfo {volume} {91}},\ \bibinfo {pages} {1505} (\bibinfo
  {year} {1953})}\BibitemShut {NoStop}%
\end{thebibliography}%

\end{document}